\documentclass{aa}
\usepackage{natbib}
\bibpunct{(}{)}{;}{a}{}{,} 

\usepackage{xcolor,colortbl}
\usepackage{wasysym}
\definecolor{MyRed}{rgb}{0.9,0.0,0.0} 
\definecolor{MyMediumBlue}{rgb}{0.7,0.72,1.0} 
\definecolor{MyGreen}{rgb}{0.0,0.9,0.5} 
\definecolor{MyDarkGreen}{rgb}{0.0,0.5,0.2} 
\definecolor{DarkRed}{rgb}{0.5,0.0,0.0} 
\definecolor{LightGrey}{rgb}{0.5,0.5,0.5}
\definecolor{sjblue}{rgb}{0.0, 0.3, 1.0}
\definecolor{sjrmblue}{rgb}{0.47, 0.62, 0.8}
\definecolor{Henrik}{rgb}{0.2,0.4,0.95}
\definecolor{HenrikII}{rgb}{0.8,0.8,0.1}
\definecolor{HenrikIII}{rgb}{0.5,0.9,0.7}

\newcommand{\hec}[1]{{#1}}
\newcommand{\hea}[1]{{#1}}

\usepackage{graphicx}
\usepackage{txfonts}

\begin{document}

   \title{The Sun at millimeter wavelengths}
  \subtitle{III. Impact of the spatial resolution on solar ALMA observations}

   \author{Henrik Eklund \inst{1,2}
    \and
    Sven Wedemeyer \inst{1,2}
    \and 
    Miko\l{}aj Szydlarski \inst{1,2}
    \and
    Shahin Jafarzadeh  \inst{1,2}
  }
  \authorrunning{Eklund {et~al.}}
   \institute{Rosseland Centre for Solar  Physics, University of Oslo, Postboks 1029 Blindern, N-0315 Oslo, Norway
            \and
            Institute of  Theoretical Astrophysics, University of Oslo, Postboks 1029 Blindern, N-0315 Oslo, Norway \\
            \email{henrik.eklund@astro.uio.no}
} 
 

   \date{Received --- ; accepted --- }

\abstract{Interferometric observations of the Sun with the Atacama Large Millimeter/sub-millimeter Array (ALMA) provide valuable diagnostic tools for studying the small-scale dynamics of the solar atmosphere. 
}
{\hec{The aims are to perform e}stimations of the observability of the small-scale dynamics as a function of spatial resolution for regions with different characteristic magnetic field topology facilitate a more robust  analysis of ALMA observations of the Sun. 
}
{A three-dimensional model of the solar atmosphere from the radiation-magnetohydrodynamic code Bifrost was used to produce high-cadence observables at \hea{millimeter and submillimeter} wavelengths. The synthetic observables for receiver bands $3$--$10$ were degraded to the angular resolution corresponding to ALMA observations with different configurations of the interferometric array from the most compact, C1, to the more extended, C7. 
The observability of the small-scale dynamics was analyzed in each case.
The analysis was thus also performed for receiver bands and resolutions that are not commissioned so far for solar observations as a means for predicting the potential of future capabilities.
}
{
The minimum resolution required to study the typical small spatial scales in the solar chromosphere depends on the characteristic properties of the target region. Here, a range from quiet Sun to enhanced network loops is considere\hea{d.} 
\hea{L}imited spatial resolution affects the observable signatures of dynamic small-scale brightening events in the form of reduced brightness temperature amplitudes, \hea{potentially leaving them undetectable,} and even shifts in the times at which the peaks occur \hea{of up to tens of seconds}. Conversion factors between the observable brightness amplitude and the original amplitude in the fully resolved simulation are provided that\hea{  can be applied to observational data in principle}, but are subject to wavelength-dependent uncertainties.  
Predictions of the typical appearance at the different combinations of receiver band, array configuration, and properties of the target region are conducted.  
}
{The simulation results demonstrate the high scientific potential that ALMA already has with the currently offered capabilities for solar observations. For the study of small-scale dynamic events, however, the spatial resolution is still crucial, and wide array configurations are preferable. In any case, it is essential to take the effects due to limited spatial resolution into account in the analysis of observational data. Finally, the further development of observing capabilities including wider array configurations and advanced imaging procedures yields a high potential for future ALMA observations of the Sun. 
}


   \maketitle

\section{Introduction}

The Atacama Large Millimeter/sub-millimeter Array (ALMA) provides great advances in observing the small-scale dynamics of the solar atmosphere for the more  direct  measurement of chromospheric temperatures as compared to other diagnostics. The \hea{millimeter (mm)} and submillimeter (sub-mm) wavelength radiation that is observable with ALMA originates from the free-free emission at chromospheric heights and is formed in local thermal equilibrium \citep[LTE; see, e.g.,][and references therein]{2016SSRv..200....1W}. The measured intensities, which are usually expressed equivalently as brightness temperatures ($T_\mathrm{b}$),
therefore have a linear relation to the local plasma temperature over the mapped atmospheric height range. 

Now, several years since the first solar ALMA observations, a number of studies of ALMA data have been reported. Some of them address the  small-scale dynamics, for example, that of magnetic field loop structures \citep{2020A&A...635A..71W}, on-disk type II spicules \citep{2021ApJ...906...82C}, 
spicules at the limb \citep{2020ApJ...888L..28S}, 
and transient brightening events with corresponding brightness temperature evolution \citep{2017ApJ...841L...5S, 2019ApJ...881...99M, 2020A&A...634A..56D, 2020A&A...638A..62N, 2020A&A...644A.152E, 2021ApJ...906...83C}, spectral sub-bands (SB) are used to estimate the optical thickness \citep{2019ApJ...875..163R}, and the dynamical structure in and around sunspots \citep{2019A&A...622A.150J} and of the oscillations in ALMA observations \citep{2020A&A...634A..86P, 2021RSPTA.37900184G, 2021RSTPA.379..174J} is studied.
\cite{2020A&A...644A.152E} systematically detected small-scale dynamics and brightening events in ALMA band~3 observations. The statistical analysis of the detected events revealed that the magnitude of the excess in brightness temperature is smaller than what is predicted based on one-dimensional \hec{(1D)} simulations \citep{2004A&A...419..747L, 2006A&A...456..713L}. 
The spatial sizes of the detected events show a distribution that increases in number toward the lower angular resolution limit ($1\arcsec.4 - 2\arcsec.1$), suggesting that there may be many more events at smaller scales that are unresolved so far. 
The spatial resolution of currently possible ALMA observations of the Sun ranges from approximately $0\arcsec.7$ and upward to several arcsec, for instance, $4\arcsec.5$ or $8\arcsec.1$ \citep{2018A&A...619L...6N, 2020A&A...638A..62N}, depending on the receiver band and array configuration. 
Chromospheric small-scale dynamic structures, with a variety of spatial and temporal scales, have also been studied with diagnostics at shorter wavelengths \hea{(e.g., optical),} allowing for a higher spatial resolution, which again depends on the size of the telescopes that are employed \citep[see, e.g.,][]{2006ApJ...648L..67V, 2006A&A...459L...9W, 2009ApJ...705..272R,2017SSRv..210..275B,2017ApJS..229....6G}.  
These have resulted in observations of small-scale structures that currently cannot be resolved with ALMA. 
As a consequence, the magnitude of the corresponding brightness temperature ($T_\mathrm{b}$) variations of the small-scale dynamics is significantly affected. 
Studies of radiative mm wavelength maps from simulations show that the $T_\mathrm{b}$ excess due to dynamic shock wave signatures is reduced due the limited spatial resolution \citep{2007A&A...471..977W}. 

\cite{2021RSPTA.37900185E} characterized the observable brightening signature from propagating shock waves in the synthetic observables at mm wavelengths from the numerical 3D magnetohydrodynamical Bifrost \citep{2011A&A...531A.154G,2016A&A...585A...4C} model, and \cite{2020A&A...644A.152E} showed that  degrading the resulting brightness temperature maps to the resolution of the observations reduces the resulting magnitude of the $T_\mathrm{b}$ excess of the brightening events. This in turn leads to a better agreement with the observational findings.
The degree of degradation of the small-scale dynamic signatures is clearly dependent on the spatial resolution, but also on the distribution of the spatial scale and on the contrasts of the target.  
The spatial resolution of an interferometer such as ALMA is mainly dependent on the physical positions of the antennas and the wavelength of the observations. 
\hea{The spatial resolution also crucially affects the identification of oscillatory signals in the solar atmosphere \citep{2021NatAs...5....5J, 2021Jess_LRSP}. In particular, a higher quiet-Sun energy flux (at a similar atmospheric height and frequency range)  was found by \citet{2010ApJ...723L.134B} from observations with a 1 m solar telescope as compared to observations with a 0.7~m telescope by \citet{2009A&A...508..941B}.} 
We use a \hec{state-of-the-art} Bifrost simulation \citep{2011A&A...531A.154G,2016A&A...585A...4C} here to perform estimations of how the observable small-scale dynamic signatures at mm and sub-mm wavelengths in both the spatial and temporal domain are affected by the degradation of the spatial resolution corresponding to ALMA observations.

\hea{This paper is the third in a paper series. The first paper by \citet{2020A&A...635A..71W} presented an analysis of one of the first regular band~3 observational data sets of the Sun and evaluated the limitations and quality of the data. The second paper in the series  \citep{2020A&A...644A.152E}  used the same observational data set to detect and study the evolution of dynamic small-scale brightening events. It concluded that the angular resolution is crucial to interpret the data.}
This third paper in the series is structured as follows. In Sect.~\ref{sec:methods} we present the numerical simulation that is used in the study and detail the calculation of the mm wavelength observables and degradation process to the spatial resolutions corresponding to ALMA observations.  In Sect.~\ref{sec:results} we present the resulting observables of the respective ALMA spectral bands. We also show the effects of limited spatial resolution on dynamical small-scale signatures in the spatial and temporal domains. Finally, in Sect.~\ref{sec:disc} we discuss some clarifications and comparisons with observational data, which are followed by conclusions in Sect.~\ref{sec:conc}.

\section{Methods}\label{sec:methods}

\subsection{3D magnetohydrodynamic simulations}\label{sec:methods-3Dmodel}
A solar atmospheric numerical 3D model from the radiation-magnetohydrodynamic code Bifrost \citep{2011A&A...531A.154G,2016A&A...585A...4C} was used, where nonlocal thermodynamic equilibrium (non-LTE) and hydrogen ionization in non-equilibrium were considered. The simulation box \hec{extends} $24$~Mm in horizontal directions ($x, y$) and $17$~Mm in vertical direction ($z$). In each horizontal direction \hec{there are} 504 cell with a uniform grid spacing of $48$~km (corresponding to an angular size of approximately $0.066$ arcsec), while in the vertical direction were $496$ cells with a grid spacing between $19-100$~km. At chromospheric heights, most relevant to this work, the vertical grid spacing \hec{is} $20$~km. The horizontal directions \hec{has} periodic boundary conditions. The bottom boundary was located at a depth of $2.5$~Mm below the solar photosphere. The total duration of the simulation sequence considered here was approximately $\text{one}$~hour with an output cadence of $1$~s to match the (currently) highest cadence of ALMA observations in solar mode. The first $200$~s of the simulation were excluded from the analysis to ensure that a quasi-equilibrium state was reached. 
The simulation represents quiet-Sun conditions, but also \hec{includes} a magnetic field region of opposite polarity in the center, in which the polarity patches \hec{are} approximately $8$~Mm apart \citep{2016A&A...585A...4C}. At photospheric heights, the overall unsigned magnetic field strength \hec{has} a value of $50$~G.
The same model was used in \citet{2020A&A...635A..71W} to study the diagnostic  potential of ALMA observations of the Sun and in \citet{2021RSPTA.37900185E} to determine the underlying physical conditions and mm wavelength signatures due to propagating shock waves. The same model setup, but with a lower cadence, was also used in \cite{2015A&A...575A..15L, 2017A&A...601A..43L} to analyze chromospheric diagnostics at mm and sub-mm wavelengths.
\hea{The magnetic topology of the simulation from the photosphere to the chromosphere and transition region} is illustrated in Figure~1 of \citet{2017ApJS..229...11J}, where the authors show that the height of magnetic canopies (i.e., when the field lines bend over from nearly vertical \hea{at low altitudes} to more horizontal configurations \hea{at higher altitudes}; \citealt{1990A&A...234..519S, 2009SSRv..144..317W}) throughout the solar atmosphere depends on the strength of the magnetic fields at their footpoints in the low photosphere.

\subsection{Brightness temperature maps at mm wavelengths corresponding to ALMA observations}\label{sec:method:rad_transfer}
The intensities at mm wavelenghts are obtained by solving the radiative transfer equation column-wise throughout the FOV, using the  advanced radiative transfer (ART) code~\citep{art_2021}\footnote{\url{https://doi.org/10.5281/zenodo.4604825}}. LTE conditions in the formation of the radiation are assumed, but relevant sources of opacity are included in detail by the code.

The intensities were calculated at frequencies corresponding to the (spectral) receiver bands of ALMA, with ten frequencies \hea{equidistantly distributed across the spectral range of each  receiver band. The spectral setup of the receiver bands is presented in Table~\ref{tab:frequencies} (see Table~\ref{tab:frequencies_appendix} for the specific frequencies that were used in the calculations for each receiver band).  
Average values of the intensities at all the ten frequencies were used in the analysis because they are a good approximation for the continuum maps derived from actual observations.}
Receiver bands $3$, $5$, $6,$ and $7$ are currently offered for solar ALMA observations. The specified wavelengths we used here are from observational cycle 8. For the remaining spectral bands ($4$, $8$, $9,$ and $10$), the central frequencies were taken to be those of the standard continuum exemplary values (Table 6.1 in the technical handbook for ALMA cycle~8\footnote{\url{https://almascience.eso.org/documents-and-tools/cycle8/alma-technical-handbook}}; \citealt{ALMA_Tech_Hand_8.3}). 
The \hea{spectral setup of these bands is symmetric around the corresponding central frequencies}
with a separation of $5$~GHz in bands $4$ and $8$ and $2$~GHz in bands $9$ and $10$ (\hea{see} Table~\ref{tab:frequencies_appendix}). If the remaining receiver bands were considered for solar observations, the resulting exact frequencies might differ slightly from what we used here, but the results still serve as a good estimate for what to expect from such potential observations.

At the mm and sub-mm wavelengths of the ALMA receiver bands, the Rayleigh-Jeans approximation holds  
and the radiative intensities ($I_\lambda$) can be translated into brightness temperatures ($T_\mathrm{b}$) through the relation
\begin{equation}
    T_\mathrm{b} = \frac{\lambda^4}{2k_\mathrm{B}c}I_\lambda
,\end{equation}
where $\lambda$ is the wavelength, $k_\mathrm{B}$ is the Boltzmann constant, and $c$ is the speed of light of the radiation.
\begin{table}[tp!]
\centering
\caption{\hea{Wavelengths and frequencies of the ALMA receiver bands assumed here. Receiver bands~$4$, $8$, $9,$ and $10$ (marked with a dagger) are not yet commissioned for solar observations and were assumed to be symmetrically distributed around the "standard continuum" central frequencies (see, e.g., Table~6.1 in the ALMA technical handbook \citealt{ALMA_Tech_Hand_8.3}), which are similar to those of the commissioned bands.}
}
\label{tab:frequencies}
\hspace*{-4mm}
\begin{tabular}{lcccc}
\hline
Receiver&\multicolumn{2}{c}{Wavelength [mm]}&\multicolumn{2}{c}{Frequency [GHz]}\\
band&min &max& min & max \\
\hline
3 & 2.7759 & 3.2586 & 92.0 & 108.0 \\
4$^\dagger$ & 1.9594 & 2.1883 & 137.0 & 153.0 \\ 
5 & 1.4553 & 1.5779 & 190.0 & 206.0 \\ 
6 & 1.2040 & 1.3091 & 229.0 & 249.0 \\ 
7 & 0.8454 & 0.8854 & 338.6 & 354.6 \\ 
8$^\dagger$ & 0.7259 & 0.7551 & 397.0 & 413.0 \\ 
9$^\dagger$ & 0.4441 & 0.4495 & 667.0 & 675.0 \\ 
10$^\dagger$ & 0.3442 & 0.3474 & 863.0 & 871.0 \\ 
\hline
\end{tabular}
\end{table}

\subsection{Spatial resolution of ALMA observations}
The spatial resolution of the interferometric array is determined by the lengths of the baselines, which are the distances between each pair of antennas.
The array can be reconfigured by physically moving some of the 12 m antennas, which thus changes the distribution of covered baselines.  
This is done in stages throughout the observational cycles, 
from the most compact array configuration C1 with a maximum baseline length of $0.16$~km to array configuration C10, which has a maximum baseline length of $16.2$~km. 

For a highly dynamic target such as the Sun, a high temporal cadence is required to capture the small-scale dynamic structures, which for long baselines comes with complications in the reconstruction of the data due to  atmospheric disturbances. 
So far (as of observational cycle $8$), only the most compact array configurations, C1 to C4 at band~$3$, C1 to C3 at band~$5$ and band~$6,$ and C1 to C2 at band~$7,$ are offered for solar observations. 
In this study, we also report estimations corresponding to a few more extended array configurations as a means of predicting the future potential for solar observations. We limit the study to array configuration C7, with a maximum baseline of $3.6$~km because configurations that are even more extended, if feasible at all for solar observations, would require substantial development and commissioning efforts and are therefore not likely to be realized in the near future. 

The shape of the point spread function (PSF) of an interferometric observation is mostly the result of the specific combination of frequencies, array configuration, and the target position on the sky relative to the baselines of the array. The sidelobes of the PSF are usually removed through a deconvolution technique, such as the commonly used CLEAN algorithm \citep{1974A&AS...15..417H}. The remaining main-lobe of the PSF, also referred to as the 'clean beam', represents the effective resolution element of the observation. 
This work shows the possibilities of observing small-scale dynamics from an optimistic perspective of a best-case scenario that corresponds to a perfectly sampled observation. Uncertainties coming from, the undersampled Fourier space, noise, and data reconstruction artifacts, for instance, would need to be added, but they are beyond the scope of the first step presented here (more about this in Sect.~\ref{sec:disc - imaging with ALMA}). 
Synthetic ALMA observations with specific frequencies and bandwidths of the receiver bands and sub-bands (cf.~Table~\ref{tab:frequencies}) and with different array configurations were constructed using the simobserve\footnote{Info about the simobserve tool can be found here: \url{https://casa.nrao.edu/casadocs-devel/stable/global-task-list/task_simobserve/about}} task included in the Common Astronomy Software Applications (CASA)\footnote{\url{https://casa.nrao.edu}} package. 
The clean beams were then extracted from the resulting measurement sets of the synthetic observations and were used as kernels in 2D convolution of the respective brightness temperature maps. 

\begin{figure}[t!]
\includegraphics[width=\columnwidth]{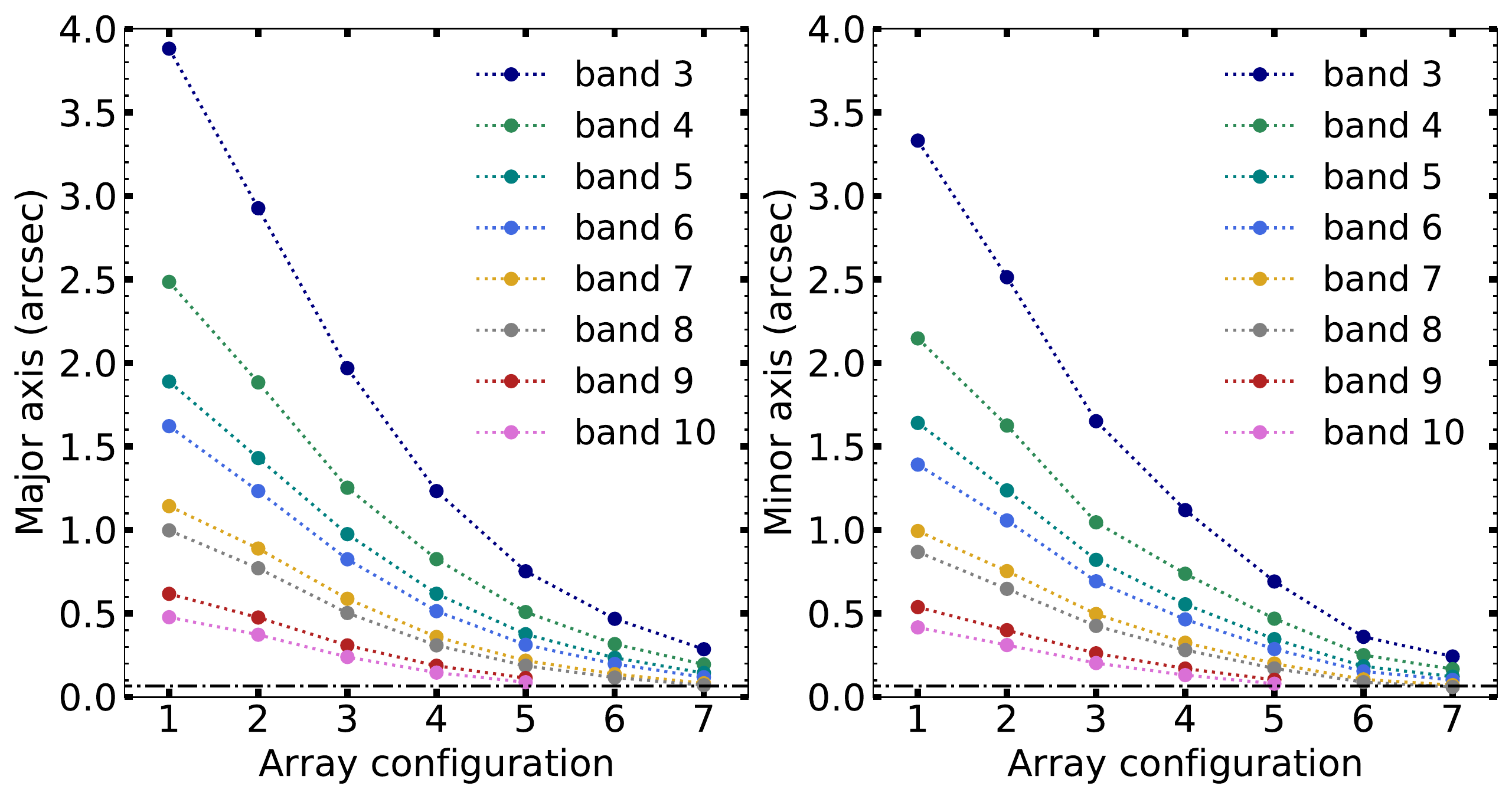}
\caption{Major and minor axes of the clean beams corresponding to the respective combination of spectral band and array configurations C1 -- C7. The values represent the FWHM along the axes of the clean beams for the full spectral width of the receiver bands at a fixed time. The horizontal dot-dashed black lines mark the limit of the cell size of $0.066$ arcsec of the original model.}
\label{fig:clean_beams}
\end{figure}

The shape of the clean beam is determined by the projected baselines on the plane of the sky perpendicular to the target source (or in other words, how the interferometric array would be seen from the target). The shape of the clean beam therefore varies with the time of the day and becomes more eccentric with increasing angular distance between the target and zenith.
The location of the Sun on the sky is therefore very important to consider when studying small-scale features. As a first approximation, we placed the Sun at an angle close to zenith, which resulted in a slightly eccentric clean beam. This choice is representative of many solar data sets found in the ALMA Science Archive\footnote{\url{https://almascience.nrao.edu/asax/}} (ASA) to this date. The minor and major axes of the clean beams for each spectral band and array configuration are given in Fig.\ref{fig:clean_beams} (numerical values are listed in Table~\ref{tab:clean_beam_parameters}).

\section{Results}\label{sec:results}

\subsection{Brightness temperature maps}

\begin{figure*}[tbh]
\sidecaption
\includegraphics[width=\textwidth]{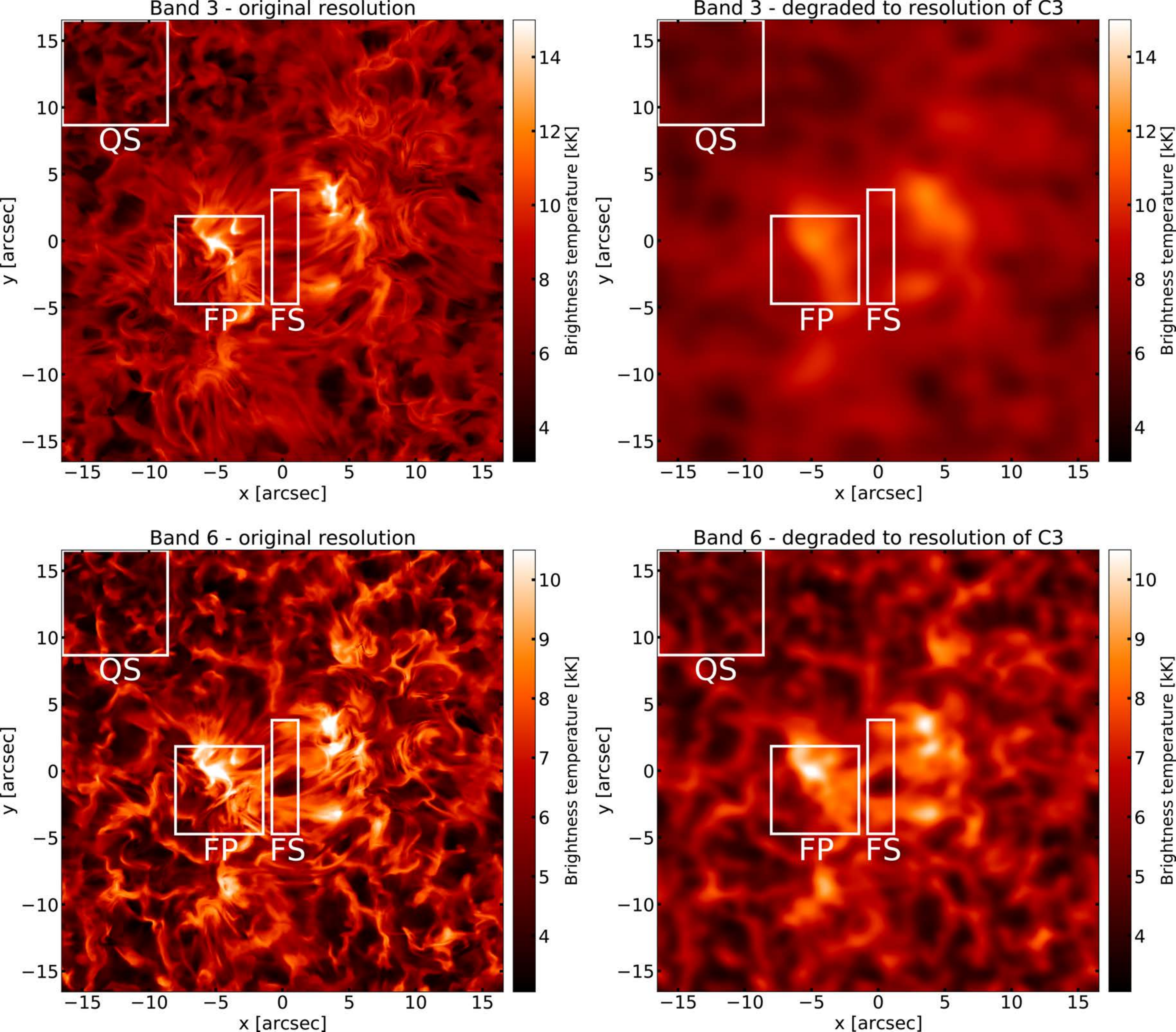}
\caption{Brightness temperature maps at selected frequencies within receiver band 3 (upper row) and band 6 (lower row) at original resolution (left column) and at degraded resolutions corresponding to observations with array configuration C3 (right column). Three selected regions are marked in the FOV: QS represents the quiet Sun, FP contains the footpoints of the prominent magnetic loops,  and FS  shows the loop structure between the two footpoints. The snapshots are taken at $t=1650$~s. The brightness temperatures at original resolution span a range of $3.1$~kK to $16.6$~kK for band 3 and $2.9$~kK to $11.3$~kK for band 6, but are capped in the figure at $15$~kK and $11$~kK, respectively, to better illustrate less bright structures.}
\label{fig:FOV}
\end{figure*}

The brightness temperature ($T_\mathrm{b}$) maps  
resulting from the radiative transfer calculations on the Bifrost model as seen from the model top (corresponding to a solar disk-center position) 
are presented in Fig.~\ref{fig:FOV} for bands~$3$ and $6$. The $T_\mathrm{b}$ maps are given at the original resolution of the model and at the degraded resolutions corresponding to those of array configuration C3 for the respective band. These bands and resolutions were chosen specifically as illustrative examples, as they represent the so-far most commonly used combinations of receiver bands and configurations for solar observations. The other combinations of receiver bands and resolutions are illustrated in \ref{sec:appendix - FOV mm maps} for reference. The brightness temperatures represent the full-band averages (see Table~\ref{tab:frequencies})\hea{.}
\hea{T}he typical contrasts and scales vary with frequency as a result of the varying formation height in the solar atmosphere (see Sect.~\ref{sec:formation_heights_full_band_continuum} for more information).
The analysis is focused on three regions with different typical characteristics, which are marked in Fig.~\ref{fig:FOV}. One inter-network quiet-Sun region (QS) and two enhanced network regions, one located straight above one of the footpoints of the magnetic loop structures (FP) and one in between the footpoints, where filamentary structures (FS) can be seen in the magnetic field loops.

\subsection{Formation heights}\label{sec:formation_heights_full_band_continuum}

\begin{figure*}[tbh]
\sidecaption
\includegraphics[width=\textwidth]{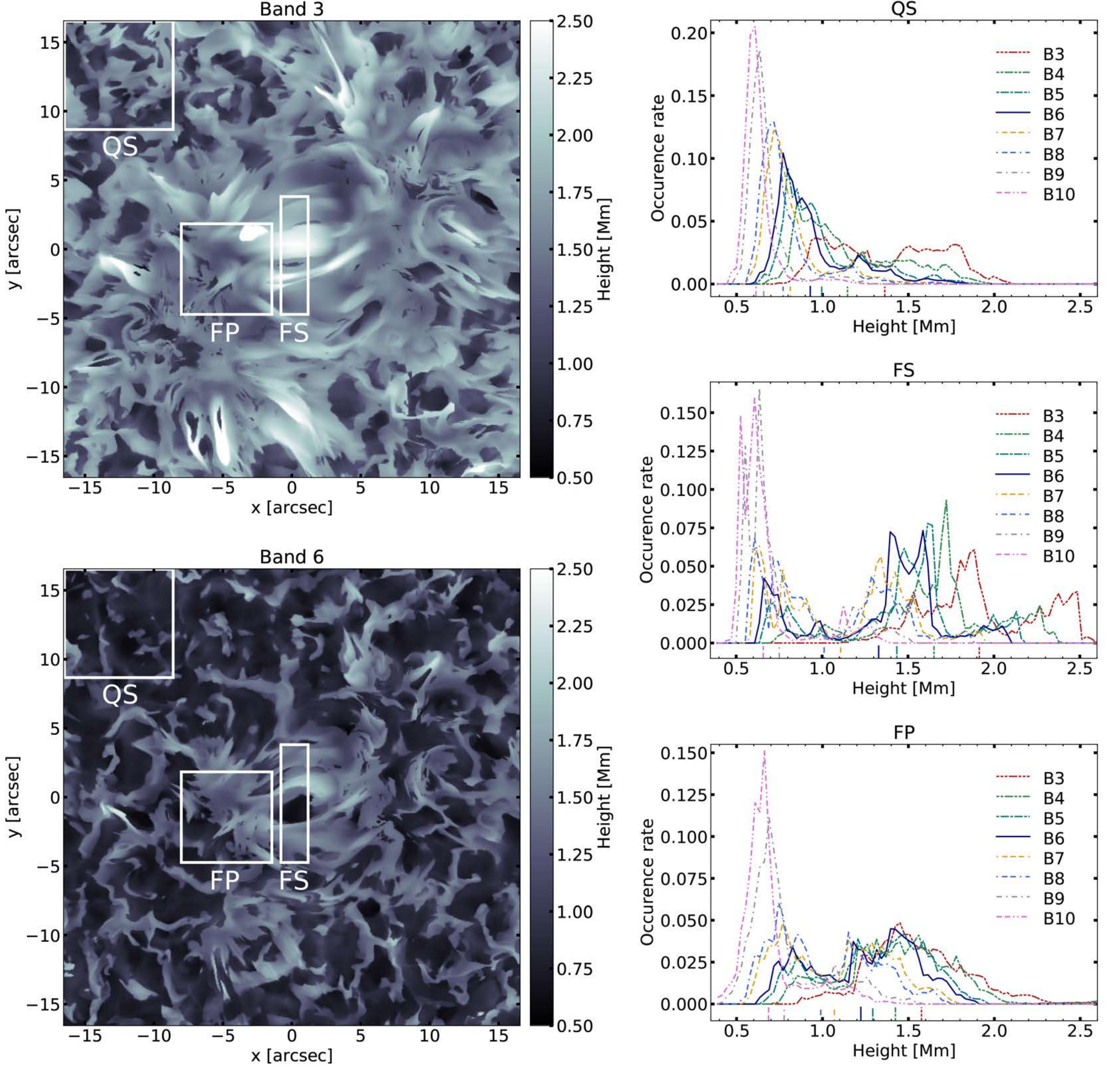}
\caption{Formation heights of the simulated continuum radiation at original resolution. Left column: Formation heights for band 3 (upper panel) and band 6 (lower panel) at  $z\,(\tau_\lambda=1)$ for the entire FOV at $t=1650$~s. 
The ranges of the FOV maps are given in the same color-scale to facilitate comparison, but the ranges span between $0.6$--$3.3$~Mm and $0.5$--$2.3$~Mm at band~3 and band~6, respectively.
Right column: Histograms of the formation heights $z\,(\tau_\lambda=1)$ of all receiver bands (3-10) in the three defined areas, QS (upper panel), FS (middle panel), and FP (lower panel). The distributions are calculated for the time range of $t=1000$--$2000$~s. The mean values of each distribution are indicated at the x-axes with matching line styles. 
\hea{The distributions are given with the same x-axis for convenience, but band~3 in the FP region in particular shows a small enhancement of up to $3$~Mm.}}
\label{fig:formation_heights_FOV}
\end{figure*}

The mm and sub-mm wavelength radiation is highly sensitive to the dynamics of the chromosphere \citep{2004A&A...419..747L, 2006A&A...456..713L}, and the formation heights varies strongly with time.
The radiation at a given wavelength originates in an extended range of heights with one or more main contributing component(s). The height of optical depth $\tau=1.0$ (at the considered observing wavelength~$\lambda$) often serves as a good approximation for the most dominant component. 
The formation heights of optical depth unity ($z(\tau_\lambda=1.0)$)  are presented in Fig.\ref{fig:formation_heights_FOV} for the entire FOV of bands 3 and 6 at one time step ($t=1650$~s), together with histogram\hea{s of the} distributions for the time range $t=1000$ -- $2000$~s for all receiver bands. 
The mean values of the formation height distribution at each band are marked at the x-axis of each corresponding region in Fig.\ref{fig:formation_heights_FOV}. 
The mean formation height of the radiation increases with wavelength.

The structures above $\sim1.5$~Mm are more extended horizontally in band 3 than in band 6 (see Fig.~\ref{fig:formation_heights_FOV}), which suggests that these structures expand with height. This is also reflected in the formation height distribution in Fig.~\ref{fig:formation_heights_FOV}, where band 3 shows significant enhancement at large heights in all the three regions compared to the other bands. This effect is most notable at heights around $1.3$ -- $1.8$~Mm in the QS region and around $2.2$ -- $2.5$~Mm in the FS region.
Most of the receiver bands show multiple components in the formation height distributions, \hea{and the height of the peaks differs between the regions.} 
For instance, in the QS region, the FOV is dominated by propagating shock waves and intermediate post-shock regions.
The radiation at mm wavelengths efficiently tracks a shock front as it propagates upward through the mid-chromosphere \citep{2021RSPTA.37900185E}. At a certain height that depends on the observed wavelength, the radiation starts  to decouple from the shock front and instead originates in lower heights in the now visible post-shock region. 
The coupling of radiation at mm wavelengths to the rapidly propagating shock fronts results in a poorer representation of the formation heights in the mid-chromosphere (cf. Fig.~3 in \citealt{2021RSPTA.37900185E}). 
\hea{This} under-representation can be seen in the QS formation height distributions in  Fig.~\ref{fig:formation_heights_FOV}, in particular, for heights of about $1.35$~Mm in band 3, $1.1$~Mm in band 6, and $1.0$~Mm in band 7. 
Bands 9 and 10 do not show a double-component distribution for the QS region, which suggests that they do not  map the upward-propagating shock fronts. These bands instead map heights in which propagating waves have not yet fully developed into shocks. 
This agrees with the typical heights seen for the onset of shock formation in the 3D numerical model  
\citep[see also, e.g.,][]{2004A&A...414.1121W}.

In the enhanced network regions, FS and \hea{FP}, 
\hea{bands} 9 and 10 mostly map the lower layers around $0.5$ -- $0.8$~Mm, whereas the lower receiver bands show larger formation heights than \hea{in} the QS region. 
With decreasing receiver band number, the magnetic loops are more efficiently mapped. 
The loops are present in a range of heights and give rise to corresponding peaks in the formation height distributions around $1.9$~Mm and $2.4$~Mm for band~$3$ and around $1.5$~Mm for band~$6$ (see Fig.~\ref{fig:formation_heights_FOV}). 
\hea{The height range below the magnetic loops (hereafter also referred to as subcanopy domain) }
is also partly sampled in band~$6$, which is reflected by the peak around $0.7$~Mm in the formation height distribution plot.

At the longest wavelengths (i.e., band 3), the formation heights in the FS region are typically larger than $1.3$~Mm, with a large peak at $1.9$~Mm and another at around $2.4$~Mm, indicating that band~$3$ maps the magnetic field loop structures at high altitudes (cf. Fig.~\ref{fig:FOV}). 
However, at shorter wavelengths (i.e., band 5 and below), an enhanced occurrence of formation heights around $0.7$~Mm is seen in the FS region, which becomes more apparent at the even shorter wavelengths of bands~$6$-$10$. This suggests that with receiver bands 5-10, we partly see through the magnetic loops, but at band 3 (similarly at band 4), the magnetic loops are to a larger extent optically thick and do not allow mapping the layers at lower heights.
The radiation in bands 9 and 10 in the FS region mostly originates at heights below the magnetic loops. This subcanopy  contribution is also picked up by lower bands, although it decreases with receiver band number and is barely present for band~3.  
The magnetic loops are substantially opaque in band~$3$  but remain partly transparent in bands~$5$~--~$7$ and to a large extent completely transparent in bands~$9$~--~$10$.
In addition, the loops are not distributed equally, so that the resulting formation height depends on the combination of the loop geometry at that location and on the opacity, which depends on the wavelength range mapped by the receiver bands.
As a result, in some locations, an opaque magnetic loop lies in the line of sight, while it is possible to see farther down into the subcanopy domain at other locations. While the specifics depend on the receiver band, there is consequently a lack of radiation with formation heights at about $1.0$~Mm for all bands.

\subsection{Brightness temperature distributions}\label{sec:results_Tb_distributions}
\begin{figure*}[tbh]
\includegraphics[width=\textwidth]{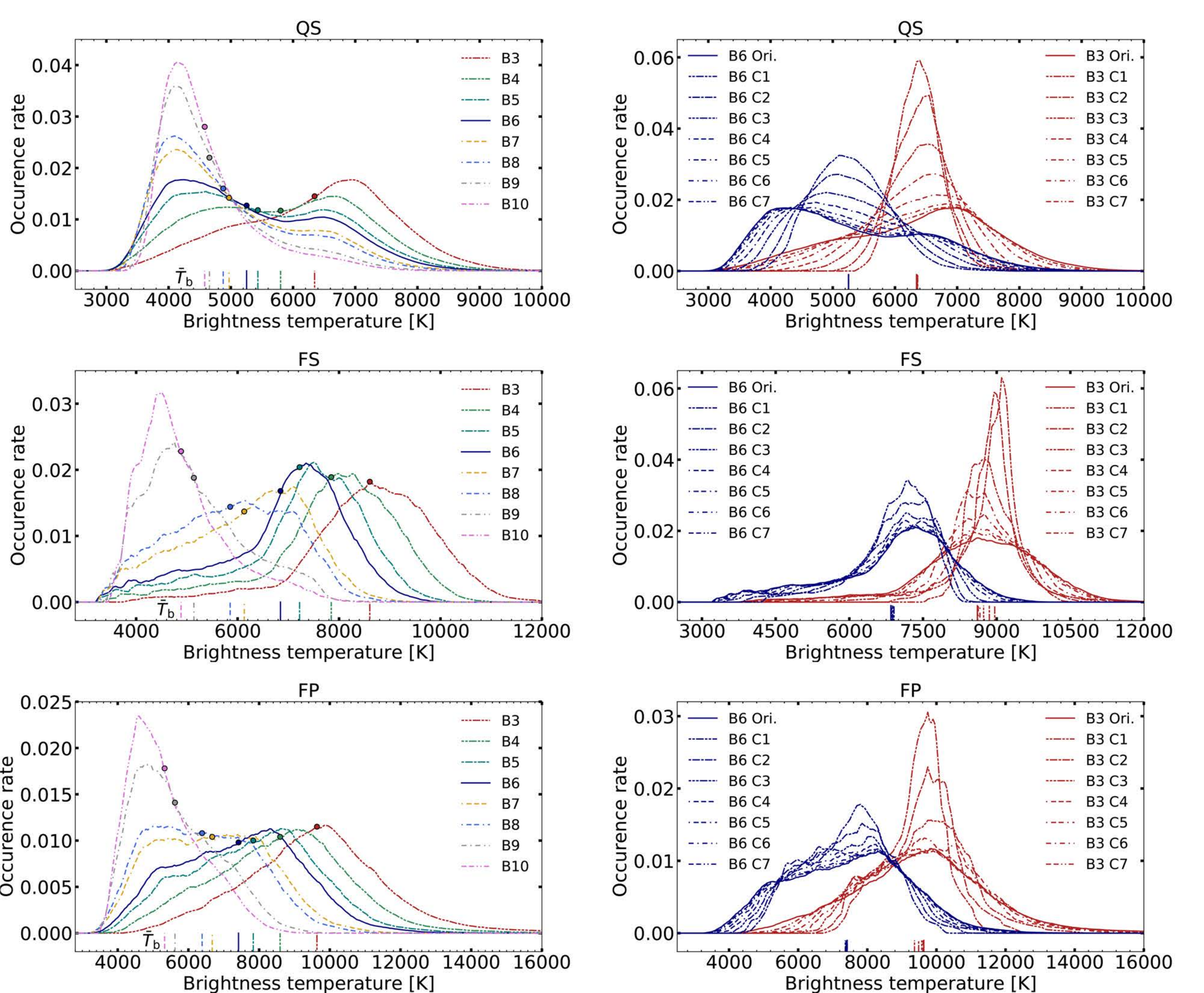}
\caption{Brightness temperature distributions corresponding to different receiver bands and array configurations. 
Left column: Brightness temperature distributions at the original resolution for the respective spectral band in each of the three different regions. Top left panel: QS. Middle left panel: FS. Bottom left panel: FP. The circles mark the mean values of each distribution, whose values are also indicated at the x-axes.
Right column: Brightness temperature distributions of bands $3$ and $6$ at the resolutions of the original model and array configurations C1 to C7. The average values of each respective distribution are marked at the x-axes. The $T_\mathrm{b}$ scales are different in each row.}
\label{fig:histograms}
\end{figure*}

In Figure~\ref{fig:histograms} we show the brightness temperature distributions for the receiver bands $3-10$  for the three selected regions, QS, FS, and FP (cf. Fig.\ref{fig:FOV}) for the time range $t=1000$ -- $2000$~s. 
There are significant differences in the $T_\mathrm{b}$ distributions between the receiver bands, which is a direct result of sampling layers at different heights at the different wavelengths (see Sect.~\ref{sec:formation_heights_full_band_continuum}).
In general, the mean value of the $T_\mathrm{b}$ distributions increases with wavelength for each of the three regions. This is expected for a monotonic increase in the average temperature in the chromosphere \citep[see, e.g.,][]{1981ApJS...45..635V}. However, there are large spans of brightness temperatures for each band and region.
The $T_\mathrm{b}$ distributions are closely related to the formation heights because the observable intensity at a given wavelength, or here equivalently, the brightness temperature, is the result of the integrated contribution function over height \hea{$z$ \citep{2021RSPTA.37900185E}, 
\begin{equation}
    T_\mathrm{b} = \int_{-\infty}^{\infty} dz ~ \chi_\nu T_\mathrm{g}(z) e^{-\tau_\nu(z)} 
,\end{equation}
where $\chi_\nu$ is the opacity, $T_\mathrm{g}$ is the gas temperature, and $\tau_\nu$ is the height dependent optical depth.
}
This connection becomes very clear for the QS region (compare the top right panel in Fig.~\ref{fig:formation_heights_FOV} with the top left panel in Fig.~\ref{fig:histograms}). 
The QS region shows the lowest mean temperatures, the FP region shows the highest temperatures, and the values for the FS region lie in between. 
The increase in $T_\mathrm{b}$ with magnetic activity is in line with recent band~3 ALMA  observations \citep{2020A&A...635A..71W}.
The shorter the wavelength (the higher the receiver  band number), the smaller the change in mean temperature between the different regions. 
The visibility of the elongated structures connected to the magnetic loops in the FS region 
becomes poorer with increasing frequency as a result of the loop geometry in combination with the mapped formation height ranges.
The magnetic field lines in the loops are largely vertical at low heights at the magnetic footpoint (in the FP region), and they gradually bend with increasing height to become practically horizontal at some height. 

\hea{The} low brightness temperatures in the middle of the FS region in the band~6 map \hea{(see Fig.~\ref{fig:FOV})} are connected to low formation heights (see Fig.~\ref{fig:formation_heights_FOV}), implying that the subcanopy domain, \hea{(i.e., the volume under the largely horizontal overlying magnetic field lines; see Sect.~\ref{sec:methods-3Dmodel})} is visible at these locations. At smaller and larger y-coordinates, however, band~6 maps higher brightness temperatures at larger heights in the magnetic loops. 
Except for a very small patch, the subcanopy domain is not visible in band~3, in contrast, because essentially the whole region is covered by magnetic loops that become opaque at heights above the subcanopy domain. Consequently, the FS region in the band~3 map is mostly filled with hotter filamentary structures.
These wavelength-dependent trends can also be seen for the remaining receiver bands, which are given in Figs.~\ref{fig:FOV_all_bands3-8}-\ref{fig:FOV_all_bands9-10} for reference.

This thus suggests that it might be possible for these magnetic field loops to study the underlying structures with short wavelengths (specifically bands~$9$~--~$10$) and to effectively study the \hea{top of the loop structures at longer wavelengths (bands~$3$ and $4$)}.
\hea{Bands}~$7$ and $8$ 
\hea{map} lower heights than bands~$5$ and $6$ and might therefore be more suitable for studying upwardly propagating waves etc.\hea{, as they enter the height range in which magnetic loops potentially interfere with the waves.} 
\hea{The effect of increasing formation height with wavelength}
\hea{and the} dependence on the magnetic filling factor of the different regions is indicated in Fig.~\ref{fig:histograms}: mean temperatures between the different regions are similar in band~$10$ ($4.58$~kK, $4.48$~kK, and $4.32$~kK in QS, FS, and FP, respectively) but show a large increase in band~$3$ from QS ($6.35$~kK) to the network region FP ($9.65$~kK).

\hea{There are a cool component and a hot component in the $T_\mathrm{b}$ distributions (Fig.~\ref{fig:histograms}), although they are more or less apparent for the individual receiver bands. 
The hot component is not visible at bands 9 and 10 because they map lower layers, but it becomes pronounced for the lower receiver bands. The peak of the distribution reaches high temperatures at about $7$~kK in band~$3$.
This effect can also be seen in the QS region in the maps for higher receiver bands, which are dominated by cool areas and show only a few, less extended hot structures (cf. Figs.~\ref{fig:FOV_all_bands3-8}-\ref{fig:FOV_all_bands9-10}). }

In Figure~\ref{fig:histograms}, the $T_\mathrm{b}$ distributions for band 3 and 6 are presented at different spatial resolutions corresponding to the original model grid spacing and to array configurations C1 to C7. 
The double-peaked QS distribution seen in the original resolution becomes less clear with decreasing resolution and tends toward a single-peaked more narrow distribution centered on the respective average value. 
In the FS and FP regions, the distributions also each tend toward a more narrow peak centered on the same average value. 

The average value of the $T_\mathrm{b}$ distribution at the original resolution of band 3 is $6346$~K in the QS region, $8606$~K in the FS region, and $9635$~K in the FP region. 
The average values of the $T_\mathrm{b}$ distributions for the degraded maps at resolutions corresponding to C1-C7 deviate by only $19$~K in the QS region, $354$~K in the FS region, and $270$~K in the FP region at most from the average values at the original resolution.
This result is expected because the spatial degradation smooths strong temperature gradients that often occur over small and thus unresolved spatial scales, resulting in a narrower $T_\mathrm{b}$ distribution around roughly the same average value. 

In band 6, the average values of the $T_\mathrm{b}$ distributions are at the original resolution $5254$~K in the QS region, $6845$~K in the FS region, and $7417$~K in the FP region. 
The shift from these average values seen at the degraded resolutions (C1~--~C7) is even less in band 6 than in band 3 (Fig.~\ref{fig:histograms}) due to the smaller angular resolution (as set by the synthesized beam). 
In the vicinity of the selected QS region, the spatial scale and $T_\mathrm{b}$ distribution change little, so that the average value therefore remains the same even if it is sampled by a beam extending outside that region. 
However, in the immediate surroundings of the FS region lie the hot magnetic footpoint areas (including the FP region), which cause in a slightly higher average $T_\mathrm{b}$ with increasingly larger sampling beam.
Similarly, the surroundings of the FP region are cooler on average than the FP region itself, resulting in a slightly lower average $T_\mathrm{b}$ with large sampling beams.

\subsection{Spatial power distributions}\label{sect:res - spatial power spectra}

\begin{figure}[t!]
\includegraphics[width=0.9\columnwidth]{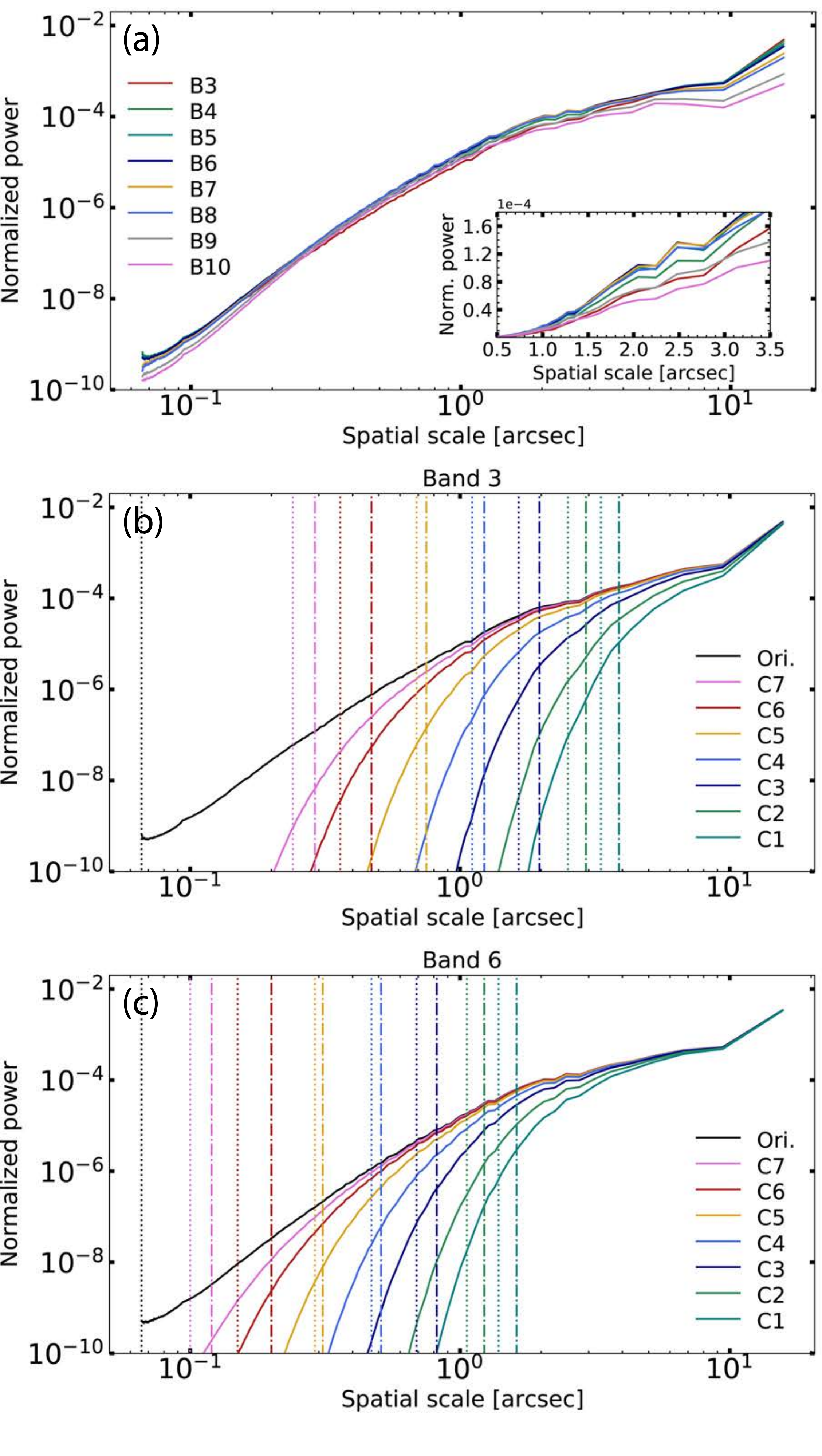}
\centering
\caption{Spatial power spectra. (a) Spatial power spectra of all bands at the original resolution, averaged over the entire FOV and between $t=1550-1750$~s. 
The inset shows a zoom-in between spatial scales of $0.5-3.5$ arcsec, with the power on a linear scale. 
(b) and (c) Spatial power spectra of bands 3 and 6 at the original resolution and the resolutions corresponding to array configurations C1-C7 for each band. The FWHMs of the minor and major axes of the corresponding clean beams are marked with the vertical dotted and dot-dashed lines, respectively. For reference, the cell size of the numerical model (at $0.066$~arcsec) is  marked by the vertical dotted black lines. }
\label{fig:degradation_example}
\end{figure}

The average spatial power spectra of the brightness temperature maps at the original resolution for the different receiver bands $3$--$10$ are shown in Fig.~\ref{fig:degradation_example}a. 
The power spectra were calculated using radially averaged 2D fast Fourier transform (FFT) on the detrended (linear) brightness temperature maps over the entire FOV and averaged over the time range $t=1550$~s~--~$1750$~s. 
The power at the larger scales ($\gtrsim 2\arcsec$) decreases with wavelength. 
This agrees with what is seen in the brightness temperature maps (Fig.~\ref{fig:FOV}), where band~6 exhibits less extended structures than band~3. This becomes even more notable by comparison with the $T_\mathrm{b}$ maps at shorter wavelengths of the upper receiver bands~$9$ and $10$ (cf. Figs.~\ref{fig:FOV_all_bands3-8}--\ref{fig:FOV_all_bands9-10}).
Moreover, at the smallest scales, a few times the cell size of the model, the power for bands~$9$ and $10$ is lower than for the lower bands.

The spatial power spectra of the degraded maps at resolutions corresponding to C1-C7 for bands~$3$ and $6$ are presented in Fig.~\ref{fig:degradation_example}b~--~c. These  power spectra are also averaged over the time range $t=1550$~s~--~$1750$~s and over the entire FOV, as for the maps at the original resolution. In each panel, the corresponding minor and major axes of the synthesized beams are shown for reference. 

The power at the degraded resolutions rapidly decreases at scales smaller than the corresponding synthesized beam.
Consequently, a resolution corresponding to configuration C3 in band~6 is at the limit for adequately studying structures at subarcsecond scales. 
However, the power spectra clearly demonstrate the improvement for observations of small-scale dynamics that is provided with the higher angular resolution corresponding to array configuration C4 and above. %
For band~3, the array configuration C5 and above would be desirable for the same reasons. 
 The reduction of the spatial power already starts at spatial scales larger than the clean-beam major axis. Depending on the scientific application, this is an important factor to consider.

\subsection{Degradation of $T_\mathrm{b}$ excess signatures of dynamic events}\label{sec:results - Tb degradation} 

\begin{figure}[tbh!]
\includegraphics[width=0.9\columnwidth]{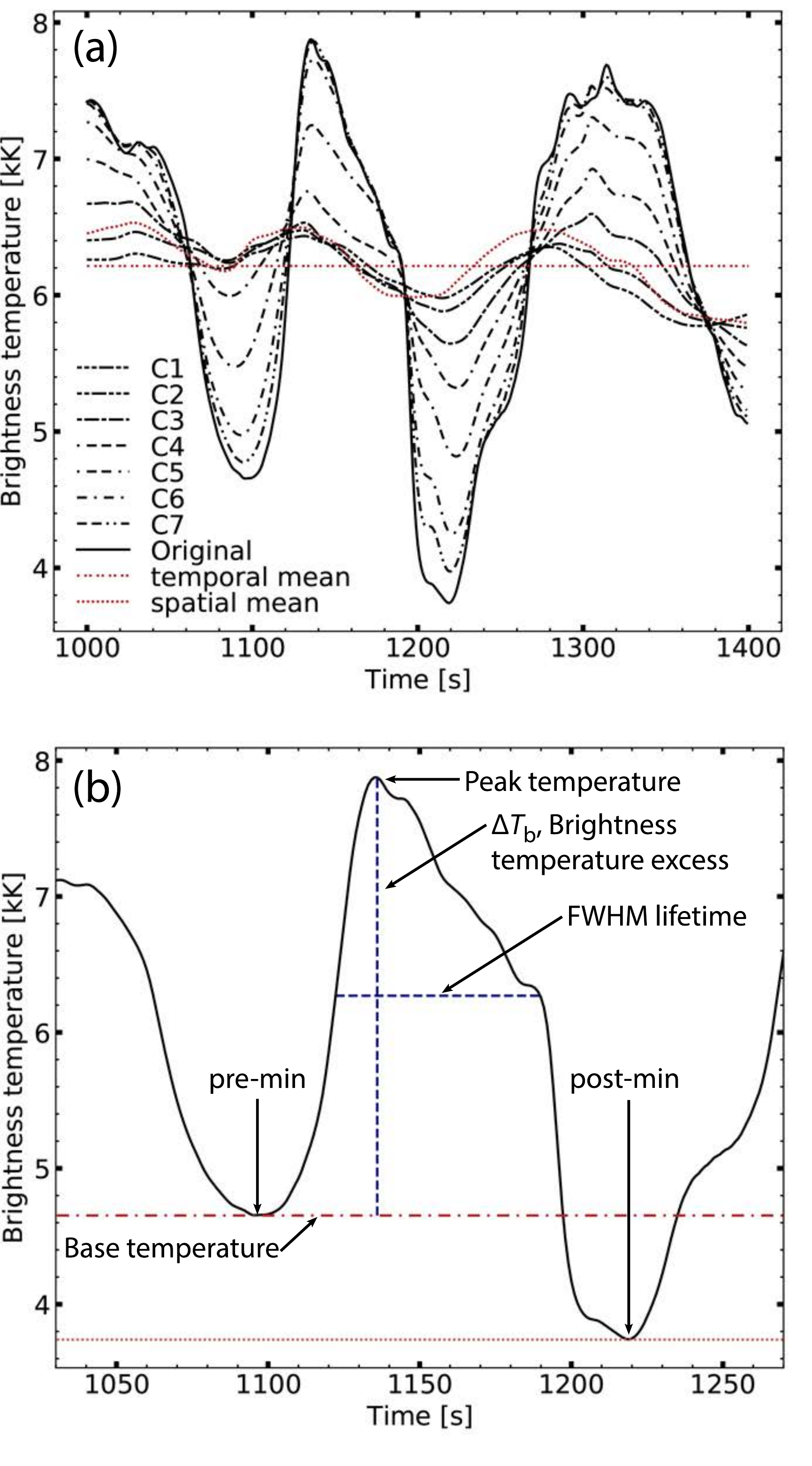}
\centering
\caption{Example of the degradation of $T_\mathrm{b}$ excess of a transient brightening event caused by limited spatial resolution and selection criteria for $T_\mathrm{b}$ excess. (a) The time evolution of the brightness temperature at $(x,y)=(-14\arcsec8, 13\arcsec1)$ in wavelengths of band~3 (cf. Table~\ref{tab:frequencies}) is given for the map with original resolution (solid line) and for the spatially degraded maps corresponding to the respective array configuration. The dotted horizontal line marks the mean $T_\mathrm{b}$ value \hea{of the displayed time series} at the original resolution, \hea{and the second dotted line shows the time evolution of the spatial mean value over a square of about $5$~arcsecondse centered on the selected location}.
(b) Illustration of the selection criteria for the dynamic events. For each peak in the temporal evolution of the brightness temperature, the local pre- and post-minima are found. The base temperature for the specific event (horizontal dash-dotted red line) is defined as the temperature of the local minima with highest temperature, and the corresponding $T_\mathrm{b}$ excess of the event (vertical dashed blue line) is the difference between the peak temperature and the base temperature. The FWHM of the peak (horizontal dashed blue line) shows the lifetime of the brightening event.}
\label{fig:Degradation_example}
\end{figure}

An example of the time evolution of the brightness temperature of at a selected position in band~3 is presented in Fig.\ref{fig:Degradation_example}a at the original resolution and at resolutions corresponding to array configurations C1-C7. This illustrates the degradation of dynamic signatures due to limited angular resolution. 
This specific example is located in the QS region 
at [$x,y$]\,=\,[$-14$\arcsec$.8, 13$\arcsec$.1$]. The adapted selection criteria for brightening events are illustrated in Fig.~\ref{fig:Degradation_example}b, where the resulting $\Delta$ $T_\mathrm{b}$ is acquired by the difference between the peak temperature and the base temperature. To be on the conservative side, the base temperature of the specific event was defined as the hottest of the local pre- and post-minima. The peak, base, and minima temperatures are thus resolution dependent and were recalculated for each resolution. 
The peak centered on $t=1140$~s \hea{(Fig.~\ref{fig:Degradation_example}a)} shows a $T_\mathrm{b}$ excess of over $3200$~K with respect to the preceding minimum at the original resolution, but only approximately $290$~K at the resolution of array configuration C3 and $240$~K at C1. This corresponds to only about $9\%$ and $7\%$, respectively, of the amplitude at the original resolution.
The degree of the amplitude degradation of the dynamic signatures is dependent on the surrounding structures, which are sampled simultaneously with the clean beam.
\hea{If a brightening event is well resolved, that is, if no other structure is mixed in within the clean beam, 
the detectable amplitude of the event will be large. 
If a brightening event is not resolved separately from the surrounding ``background'' so that  the potentially cooler ``background''  will be sampled by the same clean beam, then the resulting detectable amplitude of the brightening event will be reduced accordingly.
Because the chromosphere contains structures at scales at the same order or even smaller than the angular resolution of ALMA, the size and shape of the clean beam becomes an important factor.}
Fig.~\ref{fig:Degradation_example} shows that with larger clean-beam size (smaller array configuration number), the magnitude of the dynamic $T_\mathrm{b}$ variations deviates less from the mean value of the surrounding area.
For the same reasons that the hot brightening peaks are degraded in amplitude, the cool features would be observed as less cool (e.g., hotter) than they really are in the map at the original resolution. In this way, a small cool area in the original map can become nondetectable if there are dominant hot features in the vicinity that are sampled within the beam. For example, the cool signature around $t=1100$~s in Fig.~\ref{fig:Degradation_example}a is only weakly visible at the resolutions of C1-C3, and with temperatures close to or even above the mean value of the surrounding FOV \hea{or the temporal mean value at the selected location}, a signature like this might be easily missed or disregarded in an analysis of data at low resolution. 

An indication of the lifetime of the events is provided by the FWHM of the $\Delta T_\mathrm{b}$ peak \citep{2020A&A...644A.152E}. The lifetimes of the temperature excess typically are the same or even tend to be longer with lower resolution (see Sect.~\ref{sect:disc - Comp. with observations} for further discussion).
More importantly, the specific time at which the temperature peak occurs can shift by up to a few tens of seconds (Fig.~\ref{fig:Degradation_example}a). This causes variations in the small-scale dynamic signatures in the temporal domain. We discuss this in more detail in Sect.~\ref{sect:results - temporal power distributions}.

\begin{figure*}[tbh]
\includegraphics[width=\textwidth]{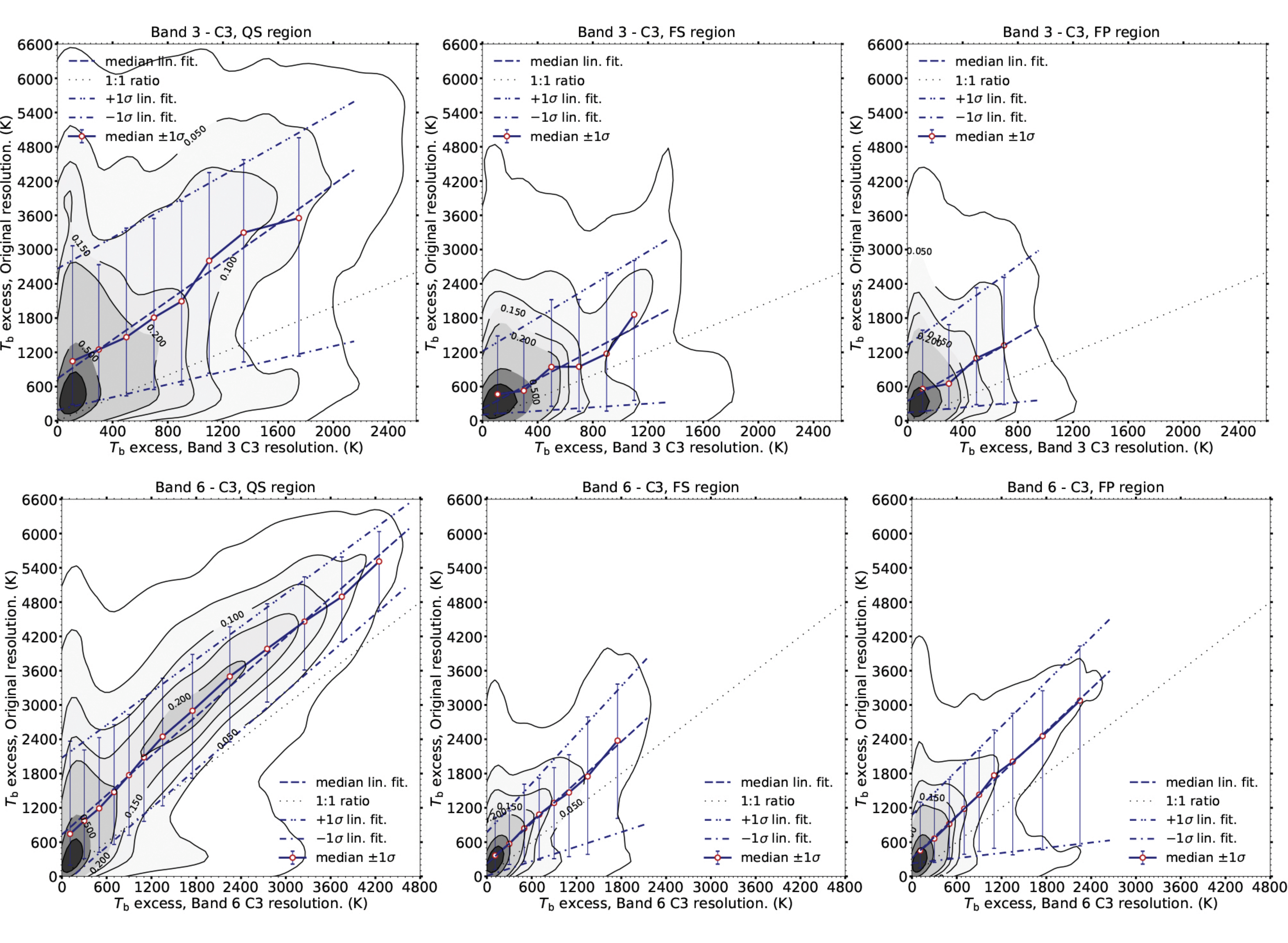}
\caption{Degradation of brightening events in band 3 (top row) and band 6 (bottom row), with the respective spatial resolution corresponding to array configuration C3. 
The $T_\mathrm{b}$ excess of brightening events at the original resolution plotted against the $T_\mathrm{b}$ excess at degraded resolution. The contour lines show the levels of 0.01, 0.05, 0.1, 0.15, 0.2, 0.5, and 0.75 of the density plot. The circles connected by the solid line show the median values (50th percentile) for the respective bin, and their error bars indicate plus and minus $1\sigma$ (e.g., 84th and 16th percentile). The dashed line marks a linear fit to the median values, and the dash-dotted lines mark linear fits to the $\pm1\sigma$ values. The dotted line marks the ratio of one-to-one for reference.}
\label{fig:hist_b3b6.pdf}
\end{figure*}

The time evolution of the brightness temperature is evaluated at all locations within the three different regions (QS, FS, and FP), between $t=1000$--$2000$~s, at the original and degraded resolution. All instances of transient brightening events with $\Delta$T$_\mathrm{b} > 20$~K are considered, \hea{well below what is reported for observed brightening events so far}. 
In Fig.~\ref{fig:hist_b3b6.pdf}, the $\Delta T_\mathrm{b}$ of all peaks at original resolution in receiver band 3 and 6 is plotted against the apparent $T_\mathrm{b}$ excess at lower angular resolutions, corresponding to the array configuration (C3) that has been used most frequently for solar observations so far. The correlations are illustrated by density distribution plots. Corresponding  plots for combinations of other spatial resolutions for array configurations C1-C7 as well as for other frequencies corresponding to receiver bands $3$--$10$ are provided in Appendix~\ref{sec:appendix_degradation_table}.

There are brightening events with larger $\Delta T_\mathrm{b}$ in the QS region than in the network regions FS or FP in both bands 3 and~6 (Fig.~\ref{fig:hist_b3b6.pdf}). Magnitudes of more than $6$~kK are seen in the QS region in both bands, but only up to about $4$~kK in the FS and FP regions. The distributions are quite broad, which results in substantial error bars for the conversion from (degraded) observed $\Delta T_\mathrm{b}$ to the true (at the original resolution) $\Delta T_\mathrm{b}$ value.
This effect is more severe at low\hea{er} resolution \hea{and is thus more evident at}  band~3 compared to\hea{ band~6}.
\hea{The} amplitudes of a majority of $\Delta T_\mathrm{b}$ peaks at degraded resolution are underestimated. This is represented by the part of the distributions above the one-to-one ratio \hea{(}indicated by the dotted lines in Fig.~\ref{fig:hist_b3b6.pdf}\hea{)}.
Some brightening events exhibit a $\Delta T_\mathrm{b}$ that is larger at the degraded resolution than at the original resolution. 
These cases are represented in the lower right \hea{corner in each} of the distribution plots (Fig.~\ref{fig:hist_b3b6.pdf}), below the one-to-one ratio. This is specifically evident in the network regions (FS and FP) \hea{at the low resolution} of band 3, where the most intense instances of $\Delta T_\mathrm{b}$ seen at the degraded resolution actually correspond to relatively low $\Delta T_\mathrm{b}$ at the original resolution. 
These $\Delta T_\mathrm{b}$ increases mostly occur in the relatively cold areas (as seen at original resolution), which host additional nearby brightening events at distances within the same  effective resolution element. These events are therefore no longer resolved individually. In these cases, a combined larger $\Delta T_\mathrm{b}$ excess is observed.
In some cases, depending on the surrounding dynamical structure, the resulting $\Delta T_\mathrm{b}$ at the degraded resolution can be on the order of up to a few thousand K, while the peak values at the original resolution only reach a few hundred K, as indicated in Fig.~\ref{fig:hist_b3b6.pdf}. 
For these reasons, the choice of specific locations for an in-depth analysis of observational data is very important.
\hea{For example, the brightest point of a brightening event will be underestimated and the surrounding cooler parts of the ``background'' will be overestimated, while the most accurate temperature is sampled somewhere in between.}

\begin{figure*}[tbh]
\includegraphics[width=0.85\textwidth]{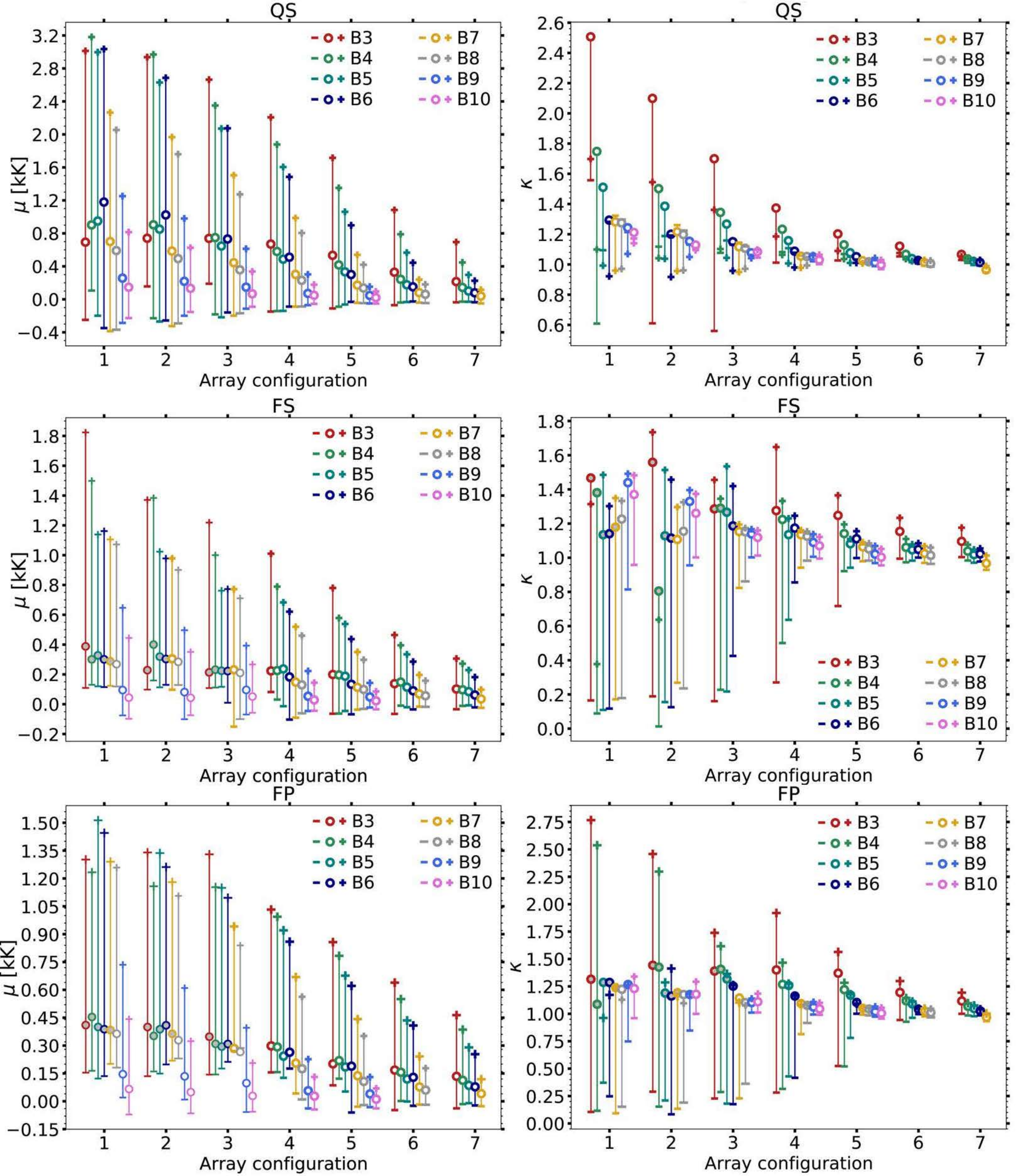}
\centering
\caption{Parameters of linear fits for Eq.~(\ref{eq:linear fits}) to the brightness temperature excess density plots (as in Fig.~\ref{fig:hist_b3b6.pdf}). All combinations of spectral bands~$3$--$10$ and array configurations C1--C7 are given for the three selected regions, QS (\hea{upper row}), FS (middle \hea{row}), and FP (\hea{bottom row}). The linear fits across the bins to the median values are indicated by the circles and to plus and minus one standard deviation are marked with plus and minus signs, respectively. The \hea{left column}, $\mu$, indicates the scalar shift of $T_\mathrm{b}$ , and the \hea{right column}, $\kappa$, indicates the slope of the linear fit. The cases in which the most intense brightening instances are dominated by apparent brightening are marked with filled gray circles.}
\label{fig:degradation_curves_parameters}
\end{figure*}

The distributions of the $\Delta$T$_\mathrm{b,deg}$ at the degraded resolution as a function of 
$\Delta$T$_\mathrm{b,org}$ at the original resolution allow us to estimate the typical degradation of $\Delta T_\mathrm{b}$ for the different types of regions considered here. The degradation factors could then in principle be applied inversely, as correction factors to observational data of similar regions. We show below, however, that these correction factors have significant error bars. 
In order to derive the typical impact on the reduced angular resolution on the simulated $\Delta T_\mathrm{b}$ amplitudes, the median and standard deviation values were calculated in the bins at degraded resolution of 
$\{0.02$ -- $0.2$, $0.2$ -- $0.4$, $0.4$ -- $0.6$, $0.6$ -- $0.8$, $0.8$ -- $1.0$, $1.0$ -- $1.2$, $1.2$ -- $1.5$, $1.5$ -- $2.0\}$~kK, 
with continuation upward of bins with a width of $0.5$~kK. In most cases, the median values and the plus/minus one standard deviation values can be reasonably well estimated by a linear fit (see Fig.~\ref{fig:hist_b3b6.pdf}). The relation between the $T_\mathrm{b}$ excess of the dynamic signatures at the original ($\Delta$T$_\mathrm{b,org}$~[kK]) and degraded ($\Delta$T$_\mathrm{b,deg}$~[kK]) resolution can thus be approximated as 
\begin{equation}\label{eq:linear fits}
    \Delta T_\mathrm{b,org} = \kappa\,(\nu,C)~ \Delta T_\mathrm{b,deg}  + \mu\,(\nu,C) 
,\end{equation}
where $\kappa$ and $\mu$ are variables that depend on the frequency $\nu$ (and thus on receiver band) and array configuration $C$ (and thus on the diameter of the synthesized telescope aperture). The values of $\kappa$ and $\mu$ for each combination of receiver bands $3$-$10$ and array configurations C1-C7 are given in Fig.~\ref{fig:degradation_curves_parameters} for the median values and plus/minus one standard deviation.
The underlying density plots are provided in Appendix~\ref{sec:appendix_degradation_table} for clarity and allow more accurate direct applications because they entail more detailed information than the approximated linear functions.
The accuracy of the translation from observed (degraded) $\Delta T_\mathrm{b}$ to true (fully resolved) $\Delta T_\mathrm{b}$ for 
dynamic brightening events in the different characteristic regions (QS, FS, and FP) at the frequencies of the spectral bands $3$ -- $10$ at the respective resolution corresponding to array configurations C1 -- C7 can thus be acquired from Fig.~\ref{fig:degradation_curves_parameters}.  
The grid cell resolution of the employed numerical model limits this study to array configurations C6 for band 8 and C5 for bands 9 and 10. 
As mentioned above, there areat certain combinations of receiver bands and array configurations, primarily corresponding to low resolution, the features with the largest $\Delta T_\mathrm{b}$ in the maps at the degraded resolution have no strong $\Delta T_\mathrm{b}$  counterpart at the original resolution. 
This is most relevant in the FS and FP regions  with the lower \hea{resolutions},
\hea{see for example the brightest events at the degraded resolution in the FS or FP regions of band~3 in Fig.~\ref{fig:hist_b3b6.pdf}.}
These events might be misinterpreted. 
These cases are marked with gray circles in Fig.~\ref{fig:degradation_curves_parameters}. 
The values in Fig.~\ref{fig:degradation_curves_parameters} are corrected for this and do not account for the distribution of the most bright apparent events.
\hea{This means that only $\Delta T_\mathrm{b}$ up to where the strongest events at the degraded resolution have a strong counterpart at the original resolution are considered in each case.} 
\hea{This can be seen in detail in the distribution plots for all combinations of receiver bands and array configurations in Appendix~\ref{sec:appendix_degradation_table}. By comparing the distributions with increasing resolution, it becomes clear at which resolution these events that appear to be most bright disappear.}

For example, for an event detected in the QS region in band~3 at the degraded resolution corresponding to array configuration C3, with an amplitude of the brightness temperature excess of $1.2$~kK, 
Fig.~\ref{fig:degradation_curves_parameters} indicates $\mu=0.74$ and $\kappa=1.7$ for the median values and for the $\pm1\sigma$; $\mu_{-\sigma}=-0.19$, $\mu_{+\sigma}=2.66$~kK, $\kappa_{-\sigma}=-0.56,$ and $\kappa_{+\sigma}= +1.36$. Inserting these values into Eq.~(\ref{eq:linear fits}) separately for the median and $\pm 1 \sigma$ , respectively, yields a span of corresponding amplitudes at the original resolution between $0.86$~kK to $4.29$~kK with a median value of $2.78$~kK, that is, a resulting estimated temperature of $T_\mathrm{b,ori}=2.78^{+1.51}_{-1.92}$~kK.

It is important to note is that as the spreads of the distributions are large already with the presented values in  Fig.~\ref{fig:degradation_curves_parameters}, which only represents the values within one standard deviation, taking into account two or even three standard deviations would in some cases give rise to more than twice the span of the distribution. The high degeneracy naturally comes from the lack of information about how well resolved the events are in the spatially degraded data.
For this reason, the distributions are not intended for performing one-to-one corrections of specific events in observational data. The results we present here instead provide an estimation of the expected uncertainties that should be accounted for when the magnitudes of brightening events in observational data are analyzed.

In conclusion, a resolution as high as possible should be used to measure the magnitudes of small-scale brightening events because it reduces the uncertainty in the conversion to the true brightness temperature excess. In particular, using array configuration C4 instead of configurations C1-C3 for band~3 observations that has been frequently used so far would greatly reduce these uncertainties.
\hea{For instance, the example above with $\Delta T_\mathrm{b} = 1.2$~kK has a corrected median value of $2.28$~kK for C4 instead of the $2.78$~kK for C3, which is $0.5$~kK smaller and corresponds to a reduction of $0.42$ times the measured $\Delta T_\mathrm{b}$.  The spread between $\pm 1 \sigma$ is reduced from $3.43$~kK at C3 to $2.52$~kK at C4, which is a difference of $0.91$~kK or $0.76$ times the measured $\Delta T_\mathrm{b}$.
}


\subsection{Temporal power distributions}\label{sect:results - temporal power distributions}

\begin{figure*}[tbh]
\includegraphics[width=0.9\textwidth]{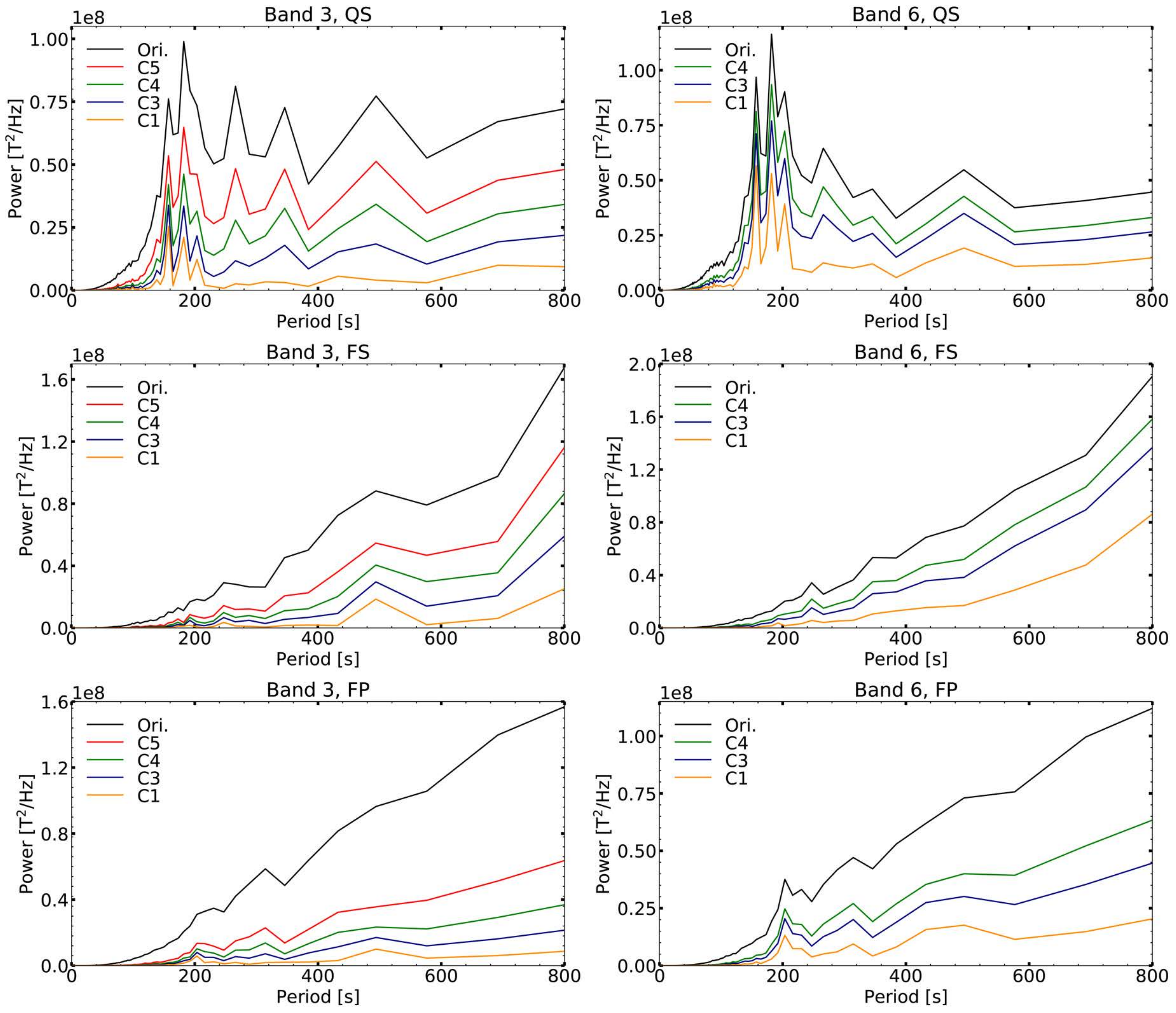}
\centering
\caption{Temporal power spectra of bands $3$ (left column) and $6$ (right column) at the three characteristically different regions (cf. Fig.~\ref{fig:FOV}). Upper row: QS region. Middle row: FS region. Bottom row: FP region. The spectra are calculated at the original resolution and at the resolutions corresponding to array configurations C1, C3, C4, and C5 for band~$3$ and C1, C3, and C4 for band~$6,$ and they represent mean values over the FOV within each region.}
\label{fig:temporal_ps36}
\end{figure*}

As indicated in the example in Fig.~\ref{fig:Degradation_example}, the appearance of the temporal $T_\mathrm{b}$ variations at  specific locations  is dependent on the surrounding spatial structures that lie within the effective resolution element. In this sense, the temporal and spatial variations are connected. The effect of limited spatial resolution is thus important when $T_\mathrm{b}$ variations in the temporal domain are studied.
As a means to inspect the effect of degraded spatial resolution on oscillatory signals, the temporal power spectra were calculated at all individual pixels over the entire time span of the simulation ($t=0$~s~--~$3460$~s) and averaged over the FOV in the respective regions. Prior to the Fourier analysis, each signal was detrended (linearly) and apodized using a tapered cosine window \citep{1978IEEEP..66...51H}.
The temporal power spectra are shown for band~$3$ and band~$6$ for the three selected regions in Fig.~\ref{fig:temporal_ps36}.

For the QS region, the power spectra for both bands~$3$ and $6$ show enhanced power at periods around $180$~s  with secondary peaks at approximately $150$~s, $200$~s, and $260$~s. There are also significant enhancements at $350$~s and $500$~s, although they are more evident in band~$3$ than in band~$6$.0
The dominating peak around $180$~s agrees well with the $3$~min oscillatory behavior seen in quiet-Sun regions at chromospheric heights in observations \citep[see, e.g.,][and references therein]{2021RSTPA.379..174J} and also simulations \citep[e.g.,][]{2004A&A...414.1121W}.  
The peak at $180$~s is also well represented in the power spectra for the degraded maps at lower spatial resolution, even at the lowest resolution of C1 in both bands, although with reduced power at all periods. 
The power of the secondary peak at $150$~s is slightly less reduced with decreasing resolution for both receiver bands as compared to the primary peak. 
As a result, the peak at 150~s appears dominant at a resolution corresponding to array configuration C1. 
The peaks at longer periods, $260$~s, $350$~s, and $500$~s, instead appear less pronounced at lower spatial resolution and are barely discernible at C1 for either bands~$3$ and $6$.

In the network regions, FS and FP, the power increases with period for bands~$3$ and $6$ due to notable temperature variations on long timescales in the numerical model in connection with the evolution of the magnetic topology.   
The power at longer periods ($>500$~s) is reduced more significantly with degraded resolution compared to those with periods shorter than 500~s.
The integrated power over all periods is highest for the QS region, followed by the FP region, and it is lowest for the FS region. The integrated power is roughly one order of magnitude lower with the degraded spatial resolution of C1 compared to the original resolution for each of the selected regions.
In the FS region, at band~$3$, \hec{there is a power enhancement} at periods between roughly $400$~s to $500$~s, which becomes more prominent at lower resolution. 
For band~$6$ in the FS region, there are no clear enhancements, but only some indications around $250$~s and $350$~s, which appear stronger with higher spatial resolution.
As indicated in Sect.~\ref{sec:formation_heights_full_band_continuum}, band~$3$ mostly samples the magnetic loop structures in the FS region, while band~$6$  also partly samples lower heights, which could suggest that the enhancement at $500$~s in band~$3$ originates from the magnetic field loops.
The magnetic footpoint region FP shows modest enhancements at short periods around $200$~s and $310$~s in both bands~$3$ and $6$ at the original resolution. As a result of the reduced power at long periods for lower spatial resolution, the enhancements appear more pronounced.

\section{Discussion}\label{sec:disc}


\subsection{Numerical model}\label{sect:disc - numerical model}
The Bifrost model provides a realistic test case because it exhibits a substantial span of spatial and temporal scales and also combines characteristic regions from quiet Sun (QS) to magnetic field footpoints (FP; Fig.~\ref{fig:FOV}). 
For these reasons, models with the same setup have been used for the study of other synthetic chromospheric observables \citep[e.g.,][]{2015ApJ...806...14P}. 
The results of this work can thus be transferred to observations of solar targets that probably show comparable conditions of the observables, as seen in any of the selected targets, such as a similar magnetic field strength and topology and similar spatial scales and contrasts.

Dynamics at  small spatial scales are often also correlated with short timescales as a direct consequence of finite velocities. In the QS region, for instance, the pattern of propagating shock fronts shows typical scales up to several arcseconds and timescales of several tens of seconds \citep{2020A&A...644A.152E, 2021RSPTA.37900185E, 2007A&A...471..977W}.
Long-term variations in the numerical model are connected to the evolution of the magnetic field topology as a result of the simulated magnetoconvection.

The magnetic field topology in the  FS region of the employed 3D model is also useful for studying the effects of limited spatial resolution on mm observations of narrow magnetic loops (see Sect.~\ref{sec:results_Tb_distributions}). In particular for the lowest resolution cases discussed here (e.g., C1-C3 of band~$3$), caution should to be taken when observational results are interpreted because the relatively large synthesised beams result in a mixing of the  magnetic field structure and its immediate surroundings. 
The same applies to all observations of features with spatial extents at or below the effective spatial resolution (see Sect.~\ref{sec:results - Tb degradation}).

\subsection{Imaging with ALMA}
\label{sec:disc - imaging with ALMA}

As a first step, the method we chose represents a best-case scenario with optimal observing conditions. 
In reality, there are additional limitations and sources of uncertainties such as
instrumental noise and the Earth's atmosphere that impact observations, especially those of small-scale dynamics. More extensive modeling is required to take these effects properly into account. This is currently limited by our incomplete knowledge of the atmosphere above the ALMA site during daytime (solar) observations. 
Furthermore, interferometric observations sample the target only at certain spatial scales and certain angles as determined by the so-called baselines of the interferometric array.  
Each baseline connects two antennas in the array and measures one spatial Fourier component of the target source.
The longest baselines determine the minimum resolvable scales (cf. Table~\ref{tab:clean_beam_parameters}), but are also more affected by atmospheric disturbances. 
Improving the spatial resolution by means of extending the interferometric array therefore comes with challenges related to the required correction of atmospheric disturbances.  
To which degree it is feasible to perform reliable solar observations in more extended array configurations needs further studies and considerations. 
Here, it is important to note that the magnitude of the corresponding uncertainties in the final images depends on the  wavelength. Typically, the longer the wavelength (the lower the receiver band number), the less affected the interferometric observation \hea{by uncertainties arising from variations in the atmosphere above the telescopes}. 
In addition, it is possible that uncertainties and artifacts are introduced to the data through post-processing procedures. 
For instance, the removal of PSF sidelobes in the initial 'dirty image' can have a very strong effect on the resulting 'clean image'. The most commonly used deconvolver algorithm for this, the CLEAN algorithm \citep{1974A&AS...15..417H}, treats the target as a collection of point sources and is suboptimal for extended targets on the Sun through its inherited nature. 
Consequently, the imaging and post-processing in general should be reviewed and adopted for solar observations in the future. 
\hea{The detailed modeling of an interferometric observation of the Sun with ALMA would require the production of artificial measurement sets, application of instrumental and atmospheric noise, and subsequent imaging. This computationally expensive approach would blend the effect of the overall angular resolution with the specifics of the employed imaging algorithm and the assumed noise properties. We therefore restrict this first study to the more straightforward convolution of synthetic mm maps with the synthesized beam, that is, the main beam of the PSF corresponding to the considered combination of receiver band and array configuration. }

While the effects described above certainly affect the quality and therefore the scientific value of the  observational data, the limited spatial resolution that we addressed in this study has a major effect on the observables and the signatures correlated to the small-scale dynamics. 
The results presented here demonstrate that it is favorable to perform observations with a spatial resolution as high as possible, which would require the commissioning and usage of wider array configurations for solar observing. 
To which degree this will result in an improvement ALMA observations of small-scale features ultimately depends on the handling of the effect of the Earth's atmosphere on the longest baselines and appropriate imaging strategies.

\subsection{Eccentricity of clean beams}\label{sect:disc - ecc clean beam}

A very important effect attached to the interferometric observations is the size and shape of the clean beam.
The size and shape can vary when antennas are removed from the measurement set in the calibration stage, for instance, which is essentially equivalent to modifying the antenna configuration. 
However, the size and shape of the clean beam unavoidably varies in time because the angle of the target on the sky changes and thus depends on the time of the day. At short timescales within an observation,  the sizes of the clean-beam axes  typically vary by about a few percent only \citep{2020A&A...635A..71W}, but this also depends on the time of the day with more rapid variations as the target recedes from zenith. 

\begin{figure}[t!]
\includegraphics[width=0.9\columnwidth]{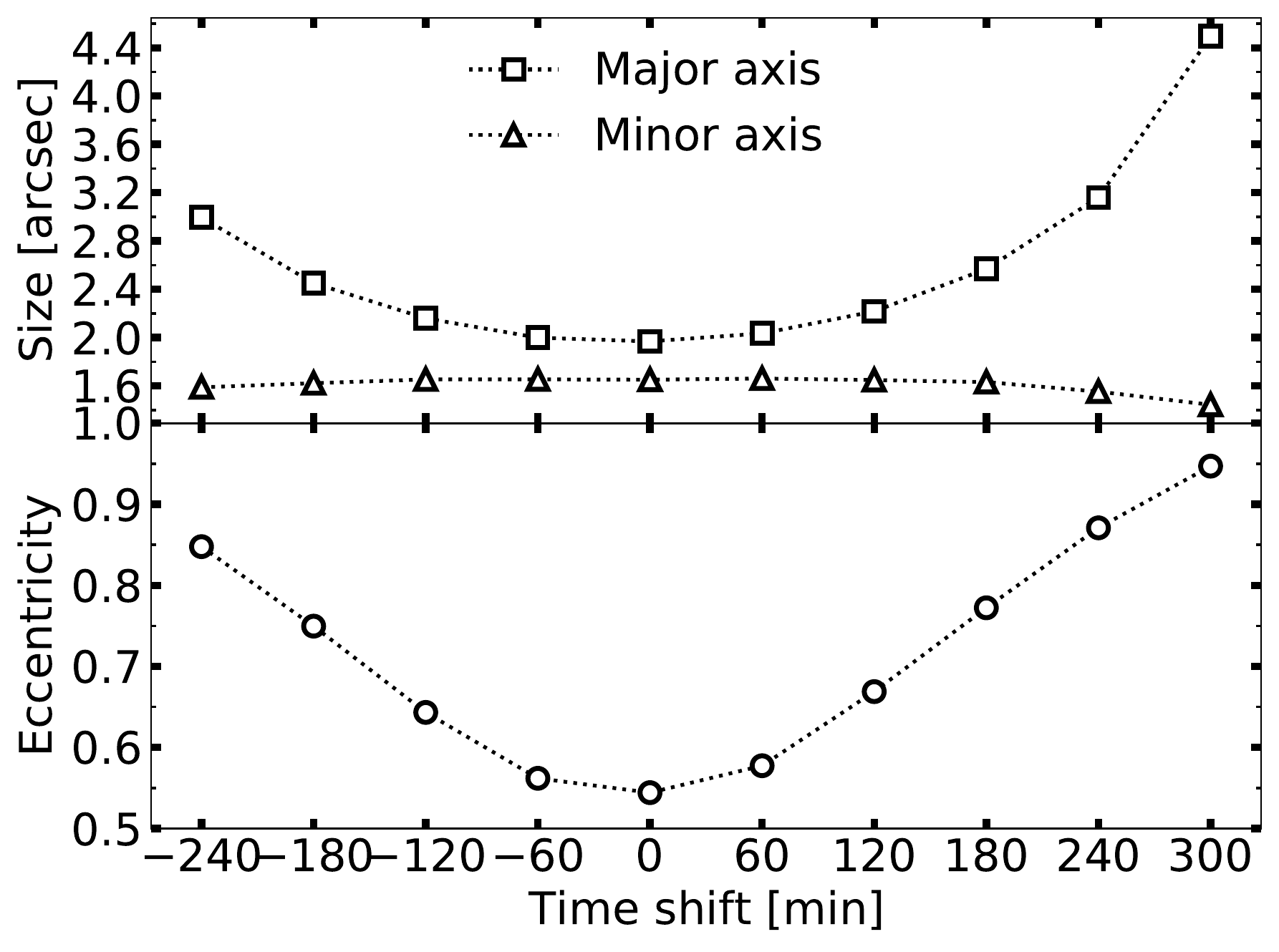}
\centering
\caption{Clean-beam size (a) and eccentricity (b) with varying time on the day. The example illustrates band~$3$ at array configuration C3, and the x-axis indicates the time shift from the standard case (close to circular clean beam) we used in our analysis.}
\label{fig:clean_beams_time_var_b3}
\end{figure}

In this study, the clean beams were kept static and slightly eccentric (cf.~Table~\ref{tab:clean_beam_parameters}), which represents more optimal conditions in which the Sun is located high on the sky. 
In Fig.~\ref{fig:clean_beams_time_var_b3}, the variations in size and eccentricity of the clean beam with the time of the day are given, with the standard case of the study as reference ($t=0$~min).
The clean-beam minor axis remains fairly constant in comparison to the major axis, which roughly doubles in size over the course of a few hours.  
The size of the clean beam plays an imperative role in the ability to resolve the small-scale structures and to the appearance of dynamic features. In addition, it is also very important to consider the shape (eccentricity) and orientation (position angle) of the clean beam in the analysis of observational data. With a highly eccentric beam shape and thus a corresponding resolution element, the small-scale structures are differently resolved in different directions. Fig.~\ref{fig:clean_beams_time_var_b3} shows that the resolution is higher in the direction along the minor axis of the clean beam. For this reason, the orientation (position angle) of the clean beam becomes increasingly important with higher eccentricity and thus for solar observations taken early or late in the day.

In this first study, the position angle was set to a constant value (of 80~degrees, Table~\ref{tab:clean_beam_parameters}) to enable one-to-one comparisons between the different cases. 
Although the resolution is relatively high along the clean-beam minor axis, the dynamics and structures on small spatial scales are still affected by larger spatial sampling in the perpendicular direction, along the major axis. In this sense, to be on the conservative side in the data analysis, the effects that an extended resolution element has on the appearance of small-scale dynamics can be estimated by matching the major axis to the beam sizes given in this work. 
We determined that an extensive study with more eccentric clean beams would be too excessive to be included in the present work, and this remains a subject for a potentially forthcoming publication.

\subsection{Comparison with observations}\label{sect:disc - Comp. with observations}

\begin{figure}[t!]
\includegraphics[width=0.95\columnwidth]{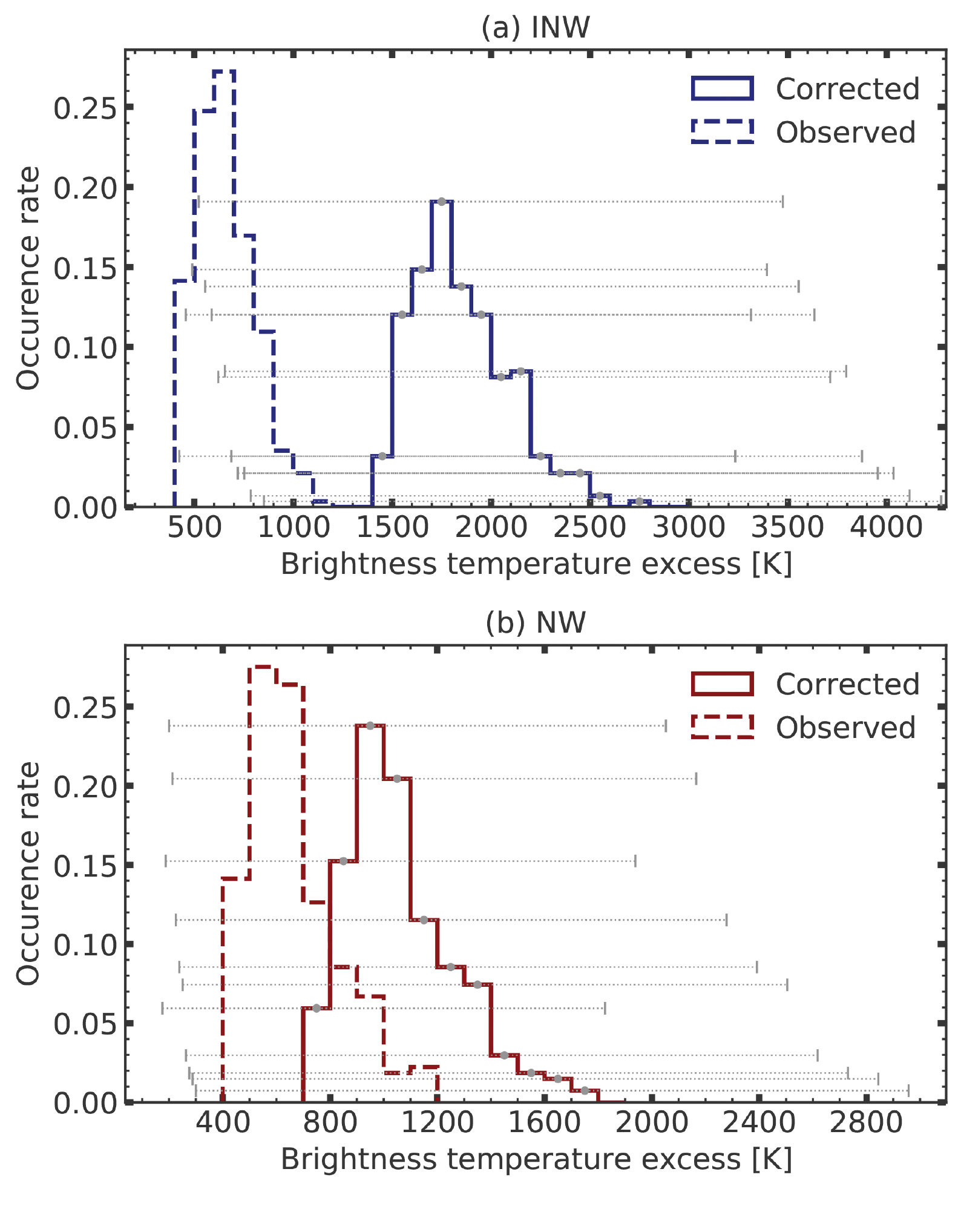}
\centering
\caption{\hea{Distributions of $\Delta T_\mathrm{b}$ of observational brightening events from \cite{2020A&A...644A.152E} with their corrected $\Delta T_\mathrm{b}$ according to  Eq.~(\ref{eq:linear fits}) and Fig.~\ref{fig:degradation_curves_parameters}. 
(a) The events observed within the INW regions (dashed blue) along with the estimated distribution using the values for QS region derived here (solid blue), with bars showing $\pm 1 \sigma$ for each bin.
(b) The observed events in the NW regions (dashed red) and the corrected distribution using the values for the FS region, with the indications of $\pm 1 \sigma$ for each bin.}
}
\label{fig:histograms_b3_obs_appli}
\end{figure}

The dependence of the observable signatures of the dynamic small-scale structure at ALMA wavelengths on the angular resolution agrees well with what is seen in studies of observational data. 
In the network regions (FS and FP), the maximum $T_\mathrm{b}$ excess of the dynamic signatures is lower than in the QS region (Fig.\ref{fig:degradation_example}). This is in line with the results from the analysis of observational band~$3$ data (taken in configuration C3) by \citet{2020A&A...644A.152E}. The reported resolution is comparable to the band~3 C3 values in this study (see Fig.~\ref{fig:clean_beams}). \citet{2020A&A...644A.152E} also reported detection of events with substantial $\Delta T_\mathrm{b}$ less frequently in areas with relatively higher magnetic field strengths, which agrees with the results presented in this study (see Sect.~\ref{sec:results - Tb degradation}).
\hea{The distributions of the observed $\Delta T_\mathrm{b}$ of the brightening events reported in \citet{2020A&A...644A.152E} are shown in Fig.~\ref{fig:histograms_b3_obs_appli} for the network (NW) and inter network (INW) regions separately (see \cite{2020A&A...644A.152E} for details).
Equation~(\ref{eq:linear fits}) is applied to 
the $\Delta T_\mathrm{b}$ of the brightening events with values $\kappa$ and $\mu$ (Fig~\ref{fig:degradation_curves_parameters}) for band~3 in array configuration C3. The QS values are used for the events observed in the INW regions and the FS values for the events observed in the NW region.
The peak of the resulting distribution of corrected $\Delta T_\mathrm{b}$ is shifted by about $1000$~K from around $650$~K to $1750$~K for the INW events and about $400$~K from around $550$~K to $950$~K for the NW events. However, the extent of the $\pm 1 \sigma$ range of the corrected $\Delta T_\mathrm{b}$ values is large. 
}
\hea{In \citet{2020A&A...644A.152E}, only events with $\Delta T_\mathrm{b} > 400$~K were considered for the detection of the brightening events, to be on the conservative side in view of the noise level in the data.}

\begin{figure}[t!]
\includegraphics[width=\columnwidth]{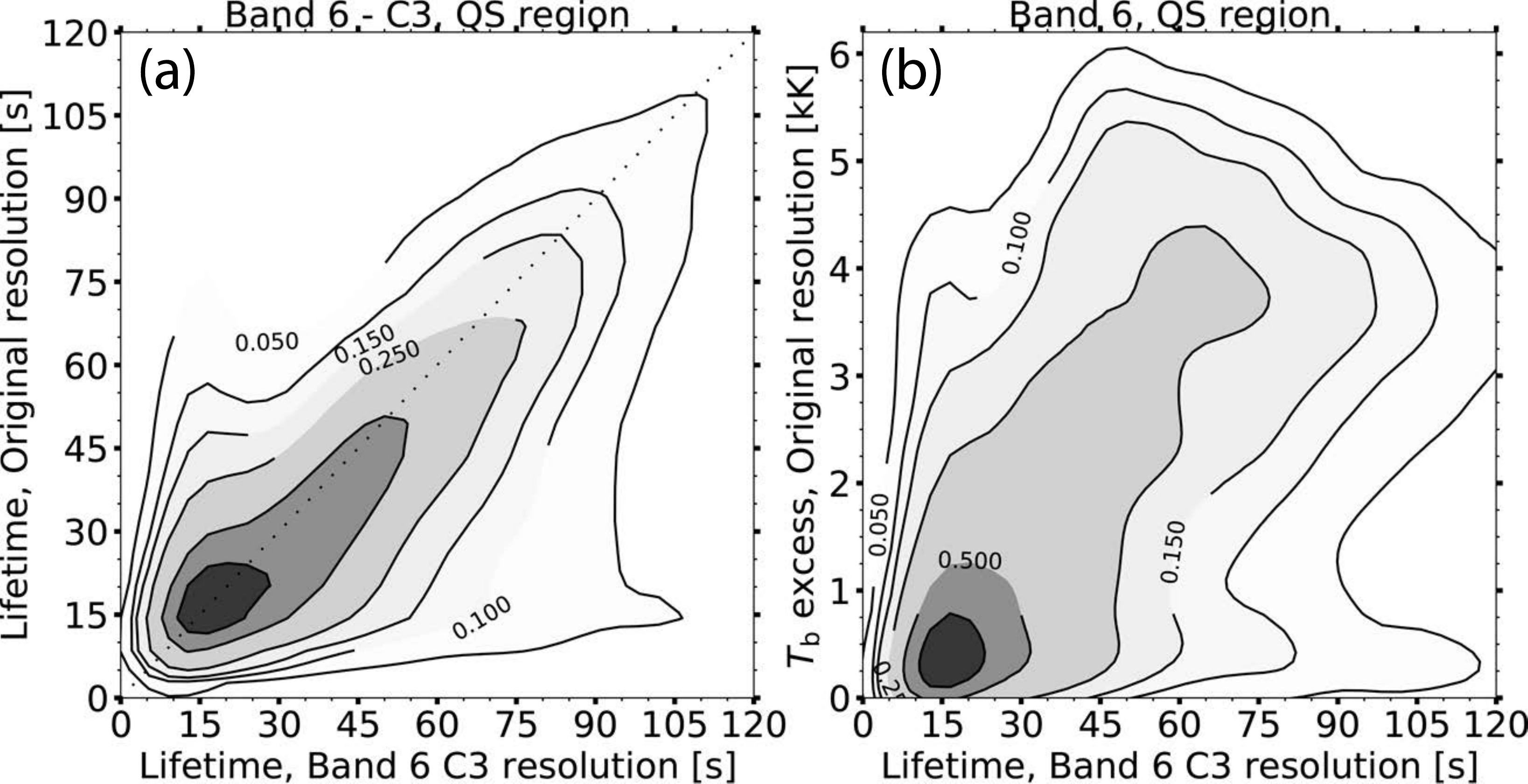}
\caption{Lifetimes of transient brightenings in the QS region at band~$6$. (a) Lifetimes at original resolution vs. lifetimes at degraded resolution corresponding to array configuration C3. (b) Brightness temperature excess at original resolution vs. lifetimes at degraded resolution corresponding to array configuration C3. The levels of the density plot, $0.05$, $0.1$, $0.15$, $0.25$, $0.5,$ and $0.75$, are marked by the contour lines.}
\label{fig:lifetime_density_plots}
\end{figure}

The lower limit of \hea{$\Delta T_\mathrm{b}$} 
\hea{for detecting brightening} signatures is dependent on the level of noise in the data.  
The effect of reduced spatial resolution presented in this study (Sect.~\ref{sec:results - Tb degradation}) could \hea{however} be used to estimate the reliability of the detected brightening events. 
At reduced resolution, the magnitudes of the $T_\mathrm{b}$ excess are suppressed (Figs.~\ref{fig:hist_b3b6.pdf}~--~\ref{fig:degradation_curves_parameters}), while there is a more linear distribution between the lifetimes of the events at degraded and original resolutions (Fig.~\ref{fig:lifetime_density_plots}a). 
For this reason, shock signatures that show a relatively small $T_\mathrm{b}$ excess in observations with low angular resolution should still show a lifetime that is in line with what is seen in the original resolution.
Detection of transient brightening events with modest $T_\mathrm{b}$ excess in comparison to the surrounding average temperature has been reported. For instance, \citet{2020A&A...638A..62N} reported events with $\Delta T_\mathrm{b} \approx  70$~K in band~$3$ data with an eccentric clean-beam size of $2.5$~arcsec$\times 4.5$~arcsec. The high eccentricity of the clean beam gives rise to larger uncertainties, as discussed in Sect.~\ref{sect:disc - ecc clean beam}. This resolution is at best (along the minor axis) comparable with array configuration C2 of band~$3$, as indicated in Fig.~\ref{fig:clean_beams}. 
Correlating these modest $\Delta T_\mathrm{b}$ signatures with their respective lifetimes could bring further insights into their origin and assist in sorting out signatures that might be due to noise or similar.
For example, for a propagating shock wave, the lifetime is several tens of seconds \citep{2021RSPTA.37900185E, 2020A&A...644A.152E} with a general trend of increasing lifetime with increasing $\Delta T_\mathrm{b}$, as indicated in Fig.~\ref{fig:lifetime_density_plots}b.

The repetitiveness of dynamic events in ALMA band~3 data of a QS region with some network patches was derived by \cite{2020A&A...644A.152E}, where a dominating period of around $200$~s, with well-represented periods up to $320$~s and declining representation up to about $430$~s, was reported. 
This agrees very well with what is seen here in the QS region of the synthetic mm maps (Fig.~\ref{fig:temporal_ps36}). 
Their analysis was restricted to the central parts of the FOV, including the most quiet parts, but similar calculations of oscillations were made for a larger FOV by \cite{2021RSTPA.379..174J}, where structures with stronger magnetic field strength were included, such as the compact magnetic magnetic field loops reported by \cite{2020A&A...635A..71W}, which may be comparable to the loops represented in the simulation model used in this study. 
The analysis of \cite{2021RSTPA.379..174J} shows that in regions with larger magnetic field strength, with overlying loops or footpoints, the resulting dominating periods are longer than in the more quiet regions.
This agrees with the periods seen in the current result of the FP region and the FS region at the resolution of C3 (Fig.~\ref{fig:temporal_ps36}).

\subsection{Comparison with previous studies of synthetic observables}\label{sect:disc - Comp. with simulations}
\hea{Synthetic brightness temperatures of radiation at mm wavelengths from 3D numerical models of the quiet Sun have previously been calculated for example by \citet{2007A&A...471..977W} and \citet{2015A&A...575A..15L}. 
A nonmagnetic model \citep{2004A&A...414.1121W} was used by \citet{2007A&A...471..977W}, who reported resulting quiet-Sun brightness temperature distributions at $1.0$~mm and $3.0$~mm with average values of $4.7$~kK and $4.9$~kK, respectively. The corresponding simulated values that we find in the current work (about $5.3$~kK at $1.2$~mm and $6.3$~kK at $3.0$~mm; see Fig.~\ref{fig:histograms}) are slightly higher, possibly as a result of the inclusion of magnetic fields in the Bifrost model.
\citet{2007A&A...471..977W} reported $T_\mathrm{b}$ distributions at $1.0$~mm and $3.0$~mm, with double-peaked characteristics as a result of sampling the cool background and hot shocks, which agrees with what we see in the current work (Fig.~\ref{fig:histograms}).
\citet{2015A&A...575A..15L} used the same publicly available Bifrost-enhanced network model 
as we used in the current work (see Sect.~\ref{sec:methods}). 
Based on one snapshot, \citeauthor{2015A&A...575A..15L}  reported average brightness temperatures over the entire FOV of $4.8$~kK at $1.0$~mm and $6.1$~kK at $3.0$~mm.
The values averaged over the ALMA bands of the current work agree well with these values, but are slightly higher because slightly longer wavelengths were included in both cases.
}

\hea{The angular resolution necessary to resolve the chromospheric shock-wave-induced mesh-like pattern was estimated by \cite{2007A&A...471..977W} by degrading a synthetic mm map at $1.0$~mm to resolutions between $0.3$~arcsec and $0.9$~arcsec. With a resolution of $0.9$~arcsec, the finest structures remain unresolved, but the larger pattern is visible.
\citet{2015A&A...575A..15L} also estimated the effect of spatial smearing by degrading the synthetic mm maps from the Bifrost snapshot to resolutions of $0.2, 0.4, 1.0,$ and $4.0$~arcsec, respectively. They concluded similarly to \cite{2007A&A...471..977W} that a resolution of up to $1.0$~arcsec is adequate for observations of the chromospheric small-scale structure. As shown in the current work, this is achieved by observations in array configuration C3 at band~6 and C5 at band~3, for instance.
}

\hea{
The average formation heights reported by \citet{2007A&A...471..977W} are about $0.77$~Mm and $0.99$~Mm for radiation at $1$~mm and $3$~mm, respectively. These values are significantly lower than the average values of the QS region in the current work, $0.93$~Mm (at $1.2$~mm) and $1.36$~Mm (at $3$~mm). One contributing factor to this might be the lack of a transition region in the model by \citet{2007A&A...471..977W}.
Previous work on dynamic \hec{1D} models by \cite{2004A&A...419..747L} shows that for some propagating shocks, radiation at $1$~mm and $3$~mm can form very close to the transition region. However, as concluded in \cite{2020A&A...644A.152E}, it is important to take the inhomogeneous structure of the 3D atmosphere into account when dynamic small-scale features are studied. 

\citet{2020ApJ...891L...8M} calculated synthetic observables at $1.2$~mm and $3.0$~mm based on a 2D numerical model including ambipolar diffusion \citep{2012ApJ...753..161M, 2017ApJ...847...36M}, for which they reported average radiation formation heights of $2.7$~Mm at $1.2$~mm and $2.8$~Mm at $3.0$~mm, which suggests that ALMA bands~3 and 6 forms at similar heights.
While the implemented modeling of relevant processes is certainly an improvement over previous modeling attempts, it is not clear whether it compensates for the restriction to two spatial dimensions. In this regard, it should be stressed that it is important to model the dynamic small-scale structure of the chromosphere in full 3D. 
Further studies are therefore needed in order to better understand the achieved degree of realism of the different models. For instance, comparison of observations at similar angular resolution at bands~3 and 6 of the same features, 
for example, magnetic field loops as reported in \cite{2020A&A...635A..71W} (where the results of this work for the FS and FP regions can be applied), 
might reveal whether the two bands show a high correlation and form at similar heights.
}

\section{Conclusion}\label{sec:conc}

Choosing the correct combination of array configuration and receiver band is essential in view of the requirements of a set science goal because it determines the achievable spatial resolution. The predictions presented here, based on \hec{state-of-the-art} simulations, clearly demonstrate that this is in particular important for the study of small-scale features. 
The amplitudes of brightness temperature signatures of transient brightening events, such as propagating shock waves, are severely reduced at low spatial resolution (Fig.~\ref{fig:hist_b3b6.pdf}), which is typically given at a compact array configuration, for example, C1-C3 for band~$3$ or C1-C2 at frequencies corresponding to up to band~$7$.
Moreover, as a result of low spatial resolution, the exact time at which a temperature peak occurs can be shifted by up to a few \hec{tens of} seconds with respect to the fully resolved case
(Fig.~\ref{fig:Degradation_example}a) because  temporal and spatial scales tend to be   coupled. 
For this reason, the observed oscillatory behavior as seen in temporal power spectra is also affected by the spatial resolution, which means that certain ranges of oscillation periods appear to be more prominently than they would be at full resolution (Fig.~\ref{fig:temporal_ps36}). 

The radiation mapped by the different ALMA receiver bands originates at different heights that range from the low chromosphere for the highest numbered receiver bands to the high chromosphere for the lowest numbered bands.  The average formation height increases with wavelength from $0.6$~Mm at band~$10$, to $0.9$~Mm at band~$6$ to $1.3$~Mm at band~$3$ in the simulated QS region.  
In the presence of magnetic field loops (FS region), the formation heights for the lower bands increase and reach an average value of up to $1.9$~Mm at band~$3,$ with small contributions from as high as $2.5$~--~$3.0$~Mm.
We note that (i)~the distribution of the formation heights of the different bands displays several components (Fig.~\ref{fig:formation_heights_FOV}) and (ii)~the formation height varies substantially in both location and time. The variations at one receiver band are on the same order or even larger than the typical differences of formation heights between the receiver bands, resulting in large overlaps that should be taken into account when relative differences between the receiver bands are studied.

Our simulations demonstrate the scientific potential of the solar observing capabilities currently provided by ALMA and how simulations can aid the in-depth analysis of observations, for instance, by providing correction factors for brightness temperature amplitudes of small-scale transient events. 
The overall conclusion is that a high spatial resolution, which is essential for studies of the dynamic small-structure of the solar chromosphere, can in principle be achieved with wider array configurations of ALMA. In order to unlock the full potential of solar observing with ALMA, further development of observing strategies and imaging procedures is required.


\section*{Acknowledgments}
This work is supported by the SolarALMA project, which has received funding from the European Research Council (ERC) under the European Union’s Horizon 2020 research and innovation programme (grant agreement No. 682462), and by the Research Council of Norway through its Centres of Excellence scheme, project number 262622.
The development of the Advanced Radiative Transfer (ART) code was supported by the PRACE Preparatory Access Type D program (proposal 2010PA3776). The support by Dr. M. Krotkiewski from USIT, University of Oslo, Norway, for the technical development of ART is gratefully acknowledged.
This paper makes use of the following ALMA data: ADS/JAO.ALMA$\#$2016.1.00423.S. ALMA is a partnership of ESO (representing its member states), NSF (USA) and NINS (Japan), together with NRC(Canada), MOST and ASIAA (Taiwan), and KASI (Republic of Korea), in co-operation with the Republic of Chile. The Joint ALMA Observatory is operated by ESO, AUI/NRAO and NAOJ. We are grateful to the many colleagues who contributed to developing the solar observing modes for ALMA and for support from the ALMA Regional Centres.

%

\bibliographystyle{aa}
\bibliography{ms}

\begin{thebibliography}{46}
\expandafter\ifx\csname natexlab\endcsname\relax\def\natexlab#1{#1}\fi

\bibitem[{{Bello Gonz{\'a}lez} {et~al.}(2009){Bello Gonz{\'a}lez}, {Flores
  Soriano}, {Kneer}, \& {Okunev}}]{2009A&A...508..941B}
{Bello Gonz{\'a}lez}, N., {Flores Soriano}, M., {Kneer}, F., \& {Okunev}, O.
  2009, \aap, 508, 941

\bibitem[{{Bello Gonz{\'a}lez} {et~al.}(2010){Bello Gonz{\'a}lez}, {Franz},
  {Mart{\'\i}nez Pillet}, {Bonet}, {Solanki}, {del Toro Iniesta}, {Schmidt},
  {Gandorfer}, {Domingo}, {Barthol}, {Berkefeld}, \&
  {Kn{\"o}lker}}]{2010ApJ...723L.134B}
{Bello Gonz{\'a}lez}, N., {Franz}, M., {Mart{\'\i}nez Pillet}, V., {et~al.}
  2010, \apjl, 723, L134

\bibitem[{{Borrero} {et~al.}(2017){Borrero}, {Jafarzadeh}, {Sch{\"u}ssler}, \&
  {Solanki}}]{2017SSRv..210..275B}
{Borrero}, J.~M., {Jafarzadeh}, S., {Sch{\"u}ssler}, M., \& {Solanki}, S.~K.
  2017, \ssr, 210, 275

\bibitem[{{Carlsson} {et~al.}(2016){Carlsson}, {Hansteen}, {Gudiksen},
  {Leenaarts}, \& {De Pontieu}}]{2016A&A...585A...4C}
{Carlsson}, M., {Hansteen}, V.~H., {Gudiksen}, B.~V., {Leenaarts}, J., \& {De
  Pontieu}, B. 2016, \aap, 585, A4

\bibitem[{{Chintzoglou} {et~al.}(2021{\natexlab{a}}){Chintzoglou}, {De
  Pontieu}, {Mart{\'\i}nez-Sykora}, {Hansteen}, {de la Cruz Rodr{\'\i}guez},
  {Szydlarski}, {Jafarzadeh}, {Wedemeyer}, {Bastian}, \& {Sainz
  Dalda}}]{2021ApJ...906...82C}
{Chintzoglou}, G., {De Pontieu}, B., {Mart{\'\i}nez-Sykora}, J., {et~al.}
  2021{\natexlab{a}}, \apj, 906, 82

\bibitem[{{Chintzoglou} {et~al.}(2021{\natexlab{b}}){Chintzoglou}, {De
  Pontieu}, {Mart{\'\i}nez-Sykora}, {Hansteen}, {de la Cruz Rodr{\'\i}guez},
  {Szydlarski}, {Jafarzadeh}, {Wedemeyer}, {Bastian}, \& {Sainz
  Dalda}}]{2021ApJ...906...83C}
{Chintzoglou}, G., {De Pontieu}, B., {Mart{\'\i}nez-Sykora}, J., {et~al.}
  2021{\natexlab{b}}, \apj, 906, 83

\bibitem[{{da Silva Santos} {et~al.}(2020){da Silva Santos}, {de la Cruz
  Rodr{\'\i}guez}, {Leenaarts}, {Chintzoglou}, {De Pontieu}, {Wedemeyer}, \&
  {Szydlarski}}]{2020A&A...634A..56D}
{da Silva Santos}, J.~M., {de la Cruz Rodr{\'\i}guez}, J., {Leenaarts}, J.,
  {et~al.} 2020, \aap, 634, A56

\bibitem[{de~la Cruz~Rodríguez {et~al.}(2021)de~la Cruz~Rodríguez,
  Szydlarski, \& Wedemeyer}]{art_2021}
de~la Cruz~Rodríguez, J., Szydlarski, M., \& Wedemeyer, S. 2021, {ART:
  Advanced (and fast!) Radiative Transfer code for Solar Physics
  (https://github.com/SolarAlma/ART).}

\bibitem[{{Eklund} {et~al.}(2021){Eklund}, {Wedemeyer}, {Snow}, {Jess},
  {Jafarzadeh}, {Grant}, {Carlsson}, \& {Szydlarski}}]{2021RSPTA.37900185E}
{Eklund}, H., {Wedemeyer}, S., {Snow}, B., {et~al.} 2021, Philosophical
  Transactions of the Royal Society of London Series A, 379, 20200185

\bibitem[{{Eklund} {et~al.}(2020){Eklund}, {Wedemeyer}, {Szydlarski},
  {Jafarzadeh}, \& {Guevara G{\'o}mez}}]{2020A&A...644A.152E}
{Eklund}, H., {Wedemeyer}, S., {Szydlarski}, M., {Jafarzadeh}, S., \& {Guevara
  G{\'o}mez}, J.~C. 2020, \aap, 644, A152

\bibitem[{{Gafeira} {et~al.}(2017){Gafeira}, {Lagg}, {Solanki}, {Jafarzadeh},
  {van Noort}, {Barthol}, {Blanco Rodr{\'\i}guez}, {del Toro Iniesta},
  {Gandorfer}, {Gizon}, {Hirzberger}, {Kn{\"o}lker}, {Orozco Su{\'a}rez},
  {Riethm{\"u}ller}, \& {Schmidt}}]{2017ApJS..229....6G}
{Gafeira}, R., {Lagg}, A., {Solanki}, S.~K., {et~al.} 2017, \apjs, 229, 6

\bibitem[{{Gudiksen} {et~al.}(2011){Gudiksen}, {Carlsson}, {Hansteen}, {Hayek},
  {Leenaarts}, \& {Mart{\'\i}nez-Sykora}}]{2011A&A...531A.154G}
{Gudiksen}, B.~V., {Carlsson}, M., {Hansteen}, V.~H., {et~al.} 2011, \aap, 531,
  A154

\bibitem[{{Guevara G{\'o}mez} {et~al.}(2021){Guevara G{\'o}mez}, {Jafarzadeh},
  {Wedemeyer}, {Szydlarski}, {Stangalini}, {Fleck}, \&
  {Keys}}]{2021RSPTA.37900184G}
{Guevara G{\'o}mez}, J.~C., {Jafarzadeh}, S., {Wedemeyer}, S., {et~al.} 2021,
  Philosophical Transactions of the Royal Society of London Series A, 379,
  20200184

\bibitem[{{Harris}(1978)}]{1978IEEEP..66...51H}
{Harris}, F.~J. 1978, IEEE Proceedings, 66, 51

\bibitem[{{H{\"o}gbom}(1974)}]{1974A&AS...15..417H}
{H{\"o}gbom}, J.~A. 1974, \aaps, 15, 417

\bibitem[{{Jafarzadeh} {et~al.}(2017){Jafarzadeh}, {Rutten}, {Solanki},
  {Wiegelmann}, {Riethm{\"u}ller}, {van Noort}, {Szydlarski}, {Blanco
  Rodr{\'\i}guez}, {Barthol}, {del Toro Iniesta}, {Gandorfer}, {Gizon},
  {Hirzberger}, {Kn{\"o}lker}, {Mart{\'\i}nez Pillet}, {Orozco Su{\'a}rez}, \&
  {Schmidt}}]{2017ApJS..229...11J}
{Jafarzadeh}, S., {Rutten}, R.~J., {Solanki}, S.~K., {et~al.} 2017, \apjs, 229,
  11

\bibitem[{{Jafarzadeh} {et~al.}(2021){Jafarzadeh}, {Wedemeyer}, {Fleck},
  {Stangalini}, {Jess}, {Morton}, {Szydlarski}, {Henriques}, {Zhu},
  {Wiegelmann}, {Guevara G{\'o}mez}, {Grant}, {Chen}, {Reardon}, \&
  {Simon}}]{2021RSTPA.379..174J}
{Jafarzadeh}, S., {Wedemeyer}, S., {Fleck}, B., {et~al.} 2021, Philosophical
  Transactions of the Royal Society A: Mathematical, 379, 20200174

\bibitem[{{Jafarzadeh} {et~al.}(2019){Jafarzadeh}, {Wedemeyer}, {Szydlarski},
  {De Pontieu}, {Rezaei}, \& {Carlsson}}]{2019A&A...622A.150J}
{Jafarzadeh}, S., {Wedemeyer}, S., {Szydlarski}, M., {et~al.} 2019, \aap, 622,
  A150

\bibitem[{{Jess} {et~al.}(2021{\natexlab{a}}){Jess}, {Jafarzadeh}, {Keys},
  {Stangalini}, \& {Verth}}]{2021Jess_LRSP}
{Jess}, D.~B., {Jafarzadeh}, S., {Keys}, P.~H., {Stangalini}, M., \& {Verth},
  G. 2021{\natexlab{a}}, lrsp, submitted

\bibitem[{{Jess} {et~al.}(2021{\natexlab{b}}){Jess}, {Snow}, {Fleck},
  {Stangalini}, \& {Jafarzadeh}}]{2021NatAs...5....5J}
{Jess}, D.~B., {Snow}, B., {Fleck}, B., {Stangalini}, M., \& {Jafarzadeh}, S.
  2021{\natexlab{b}}, Nature Astronomy, 5, 5

\bibitem[{{Loukitcheva} {et~al.}(2004){Loukitcheva}, {Solanki}, {Carlsson}, \&
  {Stein}}]{2004A&A...419..747L}
{Loukitcheva}, M., {Solanki}, S.~K., {Carlsson}, M., \& {Stein}, R.~F. 2004,
  \aap, 419, 747

\bibitem[{{Loukitcheva} {et~al.}(2015){Loukitcheva}, {Solanki}, {Carlsson}, \&
  {White}}]{2015A&A...575A..15L}
{Loukitcheva}, M., {Solanki}, S.~K., {Carlsson}, M., \& {White}, S.~M. 2015,
  \aap, 575, A15

\bibitem[{{Loukitcheva} {et~al.}(2006){Loukitcheva}, {Solanki}, \&
  {White}}]{2006A&A...456..713L}
{Loukitcheva}, M., {Solanki}, S.~K., \& {White}, S. 2006, \aap, 456, 713

\bibitem[{{Loukitcheva} {et~al.}(2017){Loukitcheva}, {White}, {Solanki},
  {Fleishman}, \& {Carlsson}}]{2017A&A...601A..43L}
{Loukitcheva}, M., {White}, S.~M., {Solanki}, S.~K., {Fleishman}, G.~D., \&
  {Carlsson}, M. 2017, \aap, 601, A43

\bibitem[{{Mart{\'\i}nez-Sykora} {et~al.}(2017){Mart{\'\i}nez-Sykora}, {De
  Pontieu}, {Carlsson}, {Hansteen}, {N{\'o}brega-Siverio}, \&
  {Gudiksen}}]{2017ApJ...847...36M}
{Mart{\'\i}nez-Sykora}, J., {De Pontieu}, B., {Carlsson}, M., {et~al.} 2017,
  \apj, 847, 36

\bibitem[{{Mart{\'\i}nez-Sykora} {et~al.}(2020){Mart{\'\i}nez-Sykora}, {De
  Pontieu}, {de la Cruz Rodriguez}, \& {Chintzoglou}}]{2020ApJ...891L...8M}
{Mart{\'\i}nez-Sykora}, J., {De Pontieu}, B., {de la Cruz Rodriguez}, J., \&
  {Chintzoglou}, G. 2020, \apjl, 891, L8

\bibitem[{{Mart{\'\i}nez-Sykora} {et~al.}(2012){Mart{\'\i}nez-Sykora}, {De
  Pontieu}, \& {Hansteen}}]{2012ApJ...753..161M}
{Mart{\'\i}nez-Sykora}, J., {De Pontieu}, B., \& {Hansteen}, V. 2012, \apj,
  753, 161

\bibitem[{{Molnar} {et~al.}(2019){Molnar}, {Reardon}, {Chai}, {Gary},
  {Uitenbroek}, {Cauzzi}, \& {Cranmer}}]{2019ApJ...881...99M}
{Molnar}, M.~E., {Reardon}, K.~P., {Chai}, Y., {et~al.} 2019, \apj, 881, 99

\bibitem[{{Nindos} {et~al.}(2018){Nindos}, {Alissandrakis}, {Bastian},
  {Patsourakos}, {De Pontieu}, {Warren}, {Ayres}, {Hudson}, {Shimizu}, {Vial},
  {Wedemeyer}, \& {Yurchyshyn}}]{2018A&A...619L...6N}
{Nindos}, A., {Alissandrakis}, C.~E., {Bastian}, T.~S., {et~al.} 2018, \aap,
  619, L6

\bibitem[{{Nindos} {et~al.}(2020){Nindos}, {Alissandrakis}, {Patsourakos}, \&
  {Bastian}}]{2020A&A...638A..62N}
{Nindos}, A., {Alissandrakis}, C.~E., {Patsourakos}, S., \& {Bastian}, T.~S.
  2020, \aap, 638, A62

\bibitem[{{Patsourakos} {et~al.}(2020){Patsourakos}, {Alissandrakis}, {Nindos},
  \& {Bastian}}]{2020A&A...634A..86P}
{Patsourakos}, S., {Alissandrakis}, C.~E., {Nindos}, A., \& {Bastian}, T.~S.
  2020, \aap, 634, A86

\bibitem[{{Pereira} {et~al.}(2015){Pereira}, {Carlsson}, {De Pontieu}, \&
  {Hansteen}}]{2015ApJ...806...14P}
{Pereira}, T. M.~D., {Carlsson}, M., {De Pontieu}, B., \& {Hansteen}, V. 2015,
  \apj, 806, 14

\bibitem[{{Remijan} {et~al.}(2020){Remijan}, {Biggs}, {Cortes}, {Dent}, {Di
  Francesco}, {Fomalont}, {Hales}, {Kameno}, {Mason}, {Philips}, {Saini},
  {Stoehr}, {Vila Vilaro}, \& {Villard}}]{ALMA_Tech_Hand_8.3}
{Remijan}, A., {Biggs}, A., {Cortes}, P., {et~al.} 2020, ALMA Doc. 8.3, ver.
  1.0, 8.3

\bibitem[{{Rodger} {et~al.}(2019){Rodger}, {Labrosse}, {Wedemeyer},
  {Szydlarski}, {Sim{\~o}es}, \& {Fletcher}}]{2019ApJ...875..163R}
{Rodger}, A.~S., {Labrosse}, N., {Wedemeyer}, S., {et~al.} 2019, \apj, 875, 163

\bibitem[{{Rouppe van der Voort} {et~al.}(2009){Rouppe van der Voort},
  {Leenaarts}, {de Pontieu}, {Carlsson}, \& {Vissers}}]{2009ApJ...705..272R}
{Rouppe van der Voort}, L., {Leenaarts}, J., {de Pontieu}, B., {Carlsson}, M.,
  \& {Vissers}, G. 2009, \apj, 705, 272

\bibitem[{{Shimojo} {et~al.}(2017){Shimojo}, {Hudson}, {White}, {Bastian}, \&
  {Iwai}}]{2017ApJ...841L...5S}
{Shimojo}, M., {Hudson}, H.~S., {White}, S.~M., {Bastian}, T.~S., \& {Iwai}, K.
  2017, \apjl, 841, L5

\bibitem[{{Shimojo} {et~al.}(2020){Shimojo}, {Kawate}, {Okamoto}, {Yokoyama},
  {Narukage}, {Sakao}, {Iwai}, {Fleishman}, \& {Shibata}}]{2020ApJ...888L..28S}
{Shimojo}, M., {Kawate}, T., {Okamoto}, T.~J., {et~al.} 2020, \apjl, 888, L28

\bibitem[{{Solanki} \& {Steiner}(1990)}]{1990A&A...234..519S}
{Solanki}, S.~K. \& {Steiner}, O. 1990, \aap, 234, 519

\bibitem[{{van Noort} \& {Rouppe van der Voort}(2006)}]{2006ApJ...648L..67V}
{van Noort}, M.~J. \& {Rouppe van der Voort}, L.~H.~M. 2006, \apjl, 648, L67

\bibitem[{{Vernazza} {et~al.}(1981){Vernazza}, {Avrett}, \&
  {Loeser}}]{1981ApJS...45..635V}
{Vernazza}, J.~E., {Avrett}, E.~H., \& {Loeser}, R. 1981, \apjs, 45, 635

\bibitem[{{Wedemeyer} {et~al.}(2016){Wedemeyer}, {Bastian}, {Braj{\v{s}}a},
  {Hudson}, {Fleishman}, {Loukitcheva}, {Fleck}, {Kontar}, {De Pontieu},
  {Yagoubov}, {Tiwari}, {Soler}, {Black}, {Antolin}, {Scullion}, {Gun{\'a}r},
  {Labrosse}, {Ludwig}, {Benz}, {White}, {Hauschildt}, {Doyle}, {Nakariakov},
  {Ayres}, {Heinzel}, {Karlicky}, {Van Doorsselaere}, {Gary}, {Alissandrakis},
  {Nindos}, {Solanki}, {Rouppe van der Voort}, {Shimojo}, {Kato},
  {Zaqarashvili}, {Perez}, {Selhorst}, \& {Barta}}]{2016SSRv..200....1W}
{Wedemeyer}, S., {Bastian}, T., {Braj{\v{s}}a}, R., {et~al.} 2016, \ssr, 200, 1

\bibitem[{{Wedemeyer} {et~al.}(2004){Wedemeyer}, {Freytag}, {Steffen},
  {Ludwig}, \& {Holweger}}]{2004A&A...414.1121W}
{Wedemeyer}, S., {Freytag}, B., {Steffen}, M., {Ludwig}, H.~G., \& {Holweger},
  H. 2004, \aap, 414, 1121

\bibitem[{{Wedemeyer} {et~al.}(2020){Wedemeyer}, {Szydlarski}, {Jafarzadeh},
  {Eklund}, {Guevara Gomez}, {Bastian}, {Fleck}, {de la Cruz Rodriguez},
  {Rodger}, \& {Carlsson}}]{2020A&A...635A..71W}
{Wedemeyer}, S., {Szydlarski}, M., {Jafarzadeh}, S., {et~al.} 2020, \aap, 635,
  A71

\bibitem[{{Wedemeyer-B{\"o}hm} {et~al.}(2009){Wedemeyer-B{\"o}hm}, {Lagg}, \&
  {Nordlund}}]{2009SSRv..144..317W}
{Wedemeyer-B{\"o}hm}, S., {Lagg}, A., \& {Nordlund}, {\r{A}}. 2009, \ssr, 144,
  317

\bibitem[{{Wedemeyer-B{\"o}hm} {et~al.}(2007){Wedemeyer-B{\"o}hm}, {Ludwig},
  {Steffen}, {Leenaarts}, \& {Freytag}}]{2007A&A...471..977W}
{Wedemeyer-B{\"o}hm}, S., {Ludwig}, H.~G., {Steffen}, M., {Leenaarts}, J., \&
  {Freytag}, B. 2007, \aap, 471, 977

\bibitem[{{W{\"o}ger} {et~al.}(2006){W{\"o}ger}, {Wedemeyer-B{\"o}hm},
  {Schmidt}, \& {von der L{\"u}he}}]{2006A&A...459L...9W}
{W{\"o}ger}, F., {Wedemeyer-B{\"o}hm}, S., {Schmidt}, W., \& {von der
  L{\"u}he}, O. 2006, \aap, 459, L9

\end{thebibliography}

\appendix
\section{Details of synthetic observations}
\label{sec:Frequencies of spectral bands}

\subsection{\hea{Specifics of the wavelengths and resolutions}}
\label{sec:appendix - frequencies - resolutions}
\hea{The receiver bands of ALMA are set up for solar observations as four sub-bands (SB), grouped into two pairs of adjacent sub-bands (i.e., for band~3, SB3.1 with SB3.2 and SB3.3 with SB3.4) at equidistant frequency separation from the central (local oscillator) frequency.}
\hea{In Table~\ref{tab:frequencies_appendix} we list the specific wavelengths and frequencies for the radiative transfer calculations. Each receiver sub-band has three frequencies, which means ten unique frequencies per receiver band, which are averaged together in the current analysis (note ten and not twelve frequencies because each pair of adjacent sub-bands shares one frequency point).}

\hea{The angular resolution is dependent on the frequency and the baselines of the interferometric array configuration. The FWHMs of the minor and major axis of the clean beams for each combination of receiver band and array configurations C1--C7 are provided in Table~\ref{tab:clean_beam_parameters}, along with the maximum baseline and the $80$~th percentile of the baselines (see \citet{ALMA_Tech_Hand_8.3}).}

\begin{table}[tbh]
\caption{\hea{Specific wavelengths and frequencies considered in the calculations of the radiative transfer}  for each of the sub-bands within each ALMA receiver band. In the notation, the sub-bands are preceded by the receiver band. $^\dagger$The receiver band is not yet commissioned for solar observations \hea{and the frequencies are symmetrically} distributed around the "standard continuum" central frequencies (see, i.e., Table~6.1 in the ALMA technical handbook \citealt{ALMA_Tech_Hand_8.3}), similar to those of the commissioned bands.
}
\label{tab:frequencies_appendix}
\hspace*{-4mm}
\begin{tabular}{lcccccc}
\hline
Sub-&\multicolumn{3}{c}{Wavelength [mm]}&\multicolumn{3}{c}{Frequency [GHz]}\\
band&min&mid &max& min & mid  & max \\
\hline                
SB3.1 & 3.1893 &3.2236& 3.2586 &  92.0 &93.0&  94.0 \\ 
SB3.2 & 3.1228 &3.1557& 3.1893 &  94.0 &95.0&  96.0 \\
SB3.3 & 2.8282 &2.8552& 2.8826 & 104.0 &105.0& 106.0 \\
SB3.4 & 2.7759 &2.8018& 2.8282 & 106.0 &107.0& 108.0 \\
 \hline
SB4.1$^\dagger$& 2.1568 & 2.1724 & 2.1883 & 137.0 & 138.0 & 139.0\\ 
SB4.2$^\dagger$& 2.1262 & 2.1414  & 2.1568 & 139.0 & 140.0 & 141.0\\ 
SB4.3$^\dagger$& 1.9854 & 1.9986 & 2.0120 & 149.0 & 150.0 & 151.0\\ 
SB4.4$^\dagger$& 1.9594 & 1.9723  & 1.9854 & 151.0 & 152.0 & 153.0\\ 
 \hline
SB5.1& 1.5614 & 1.5696 & 1.5779 & 190.0 & 191.0 & 192.0\\ 
SB5.2& 1.5453 & 1.5533 & 1.5614 & 192.0 & 193.0 & 194.0\\ 
SB5.3& 1.4696 & 1.4768 & 1.4841 & 202.0 & 203.0 & 204.0\\ 
SB5.4& 1.4553 & 1.4624 & 1.4696 & 204.0 & 205.0 & 206.0\\ 
 \hline
SB6.1 & 1.2978 &1.3034& 1.3091 & 229.0&230.0& 231.0 \\ 
SB6.2 & 1.2867 &1.2922& 1.2978 & 231.0&232.0& 233.0 \\
SB6.3 & 1.2137 &1.2187& 1.2236 & 245.0 &246.0& 247.0 \\
SB6.4 & 1.2040 &1.2088& 1.2137 & 247.0&248.0& 249.0 \\
 \hline
SB7.1& 0.8802 &0.8828& 0.8854 & 338.6 &339.6& 340.6 \\ 
SB7.2& 0.8751 &0.8776& 0.8802 & 340.6 &341.6& 342.6 \\
SB7.3& 0.8502 &0.8527& 0.8551 & 350.6 &351.6& 352.6 \\
SB7.4& 0.8454 &0.8478& 0.8502 & 352.6 &353.6& 354.6 \\
  \hline
SB8.1$^\dagger$& 0.7514 & 0.7532 & 0.7551 & 397.0 & 398.0 & 399.0\\
SB8.2$^\dagger$& 0.7476 & 0.7495 & 0.7514 & 399.0 & 400.0 & 401.0\\
SB8.3$^\dagger$& 0.7294 & 0.7312 & 0.7330 & 409.0 & 410.0 & 411.0\\
SB8.4$^\dagger$& 0.7259 & 0.7277  & 0.7294 & 411.0 & 412.0 & 413.0\\
  \hline
SB9.1$^\dagger$& 0.4481 & 0.4488  & 0.4495 & 667.0 & 668.0 & 669.0\\ 
SB9.2$^\dagger$& 0.4468 & 0.4475 & 0.4481 & 669.0 & 670.0 & 671.0\\ 
SB9.3$^\dagger$& 0.4455 & 0.4461 & 0.4468 & 671.0 & 672.0 & 673.0\\ 
SB9.4$^\dagger$& 0.4441 & 0.4448 & 0.4455& 673.0 & 674.0 & 675.0\\
 \hline
SB10.1$^\dagger$& 0.3466 & 0.3470 & 0.3474 & 863.0 & 864.0 & 865.0\\ 
SB10.2$^\dagger$& 0.3458 & 0.3462 & 0.3466 & 865.0 & 866.0 & 867.0\\ 
SB10.3$^\dagger$& 0.3450 & 0.3454 & 0.3458 & 867.0 & 868.0 & 869.0\\ 
SB10.4$^\dagger$& 0.3442 & 0.3446 & 0.3450 & 869.0 & 870.0 & 871.0\\
\hline
\end{tabular}
\end{table}

\begin{table*}[tp!]
\caption{Clean-beam parameters used in the analysis of the full spectral bands in the realistic good-case scenario with an only slightly elliptical beam. The position angle given in the fifth column is the resulting output from the simulated observations, but all the setups were given the same value of the position angle, $\phi=80\deg$ to enable one-to-one comparisons. The array configurations we used are those for cycle 5 included in the simobserve tool. The ACA array was included.  
The spectral bands that are not yet commissioned for solar observations (marked with a dagger) are assigned frequencies (cf.~Table~\ref{tab:frequencies_appendix}) according to the standard continuum values \citep[see ALMA cycle 8 technical handbook,][]{ALMA_Tech_Hand_8.3}.
The array configurations that are not yet commissioned for solar observations with the specific spectral band are marked with an asterisk.
The longest baseline of the respective configuration is given in Col. 5 and the 80th percentile baseline in Col. 6 \citep{ALMA_Tech_Hand_8.3}.}
\label{tab:clean_beam_parameters}
\centering
\begin{tabular}{llcccc}
\hline
Reciever&Array&\multicolumn{2}{c}{Clean beam} & Maximum & 80th percentile \\
band&configuration &Major axis [arcsec] & Minor axis [arcsec] & baseline [km] & of baselines [km]\\
\hline
Band 3 & C1 & 3.8809 & 3.3305 & 0.1607 & 0.1071\\
& C2 & 2.9258 & 2.5128 & 0.3137 & 0.1438\\
& C3 & 1.9682 & 1.6511 & 0.5002 & 0.2354\\
& C4 & 1.2329& 1.1191 & 0.7835 & 0.3692\\
& C5* & 0.7527 & 0.6916 & 1.3979 & 0.6238\\
& C6* & 0.4683 & 0.3605 & 2.5169 & 1.1725\\
& C7* & 0.2862 & 0.2429 & 3.6378 & 1.6731\\
\hline
Band 4$\dagger$ & C1* & 2.4849 & 2.1466 & 0.1607 & 0.1071\\
& C2* & 1.8832 & 1.6250 & 0.3137 & 0.1438\\
& C3* & 1.2527 & 1.0457 & 0.5002 & 0.2354\\
& C4* & 0.8256 & 0.7380  &0.7835 & 0.3692\\
& C5* & 0.5089 & 0.4691 &1.3979 & 0.6238\\
& C6* & 0.3171 & 0.2508 &2.5169 & 1.1725\\
& C7* & 0.1954 & 0.1673  &3.6378 & 1.6731\\
\hline
Band 5 & C1 & 1.8890 & 1.6405 & 0.1607 & 0.1071\\
& C2 & 1.4304 & 1.2370 &  0.3137 & 0.1438\\
& C3 & 0.9751 & 0.8211 &  0.5002 & 0.2354\\
& C4* & 0.6181 & 0.5551 &  0.7835 & 0.3692\\
& C5* & 0.3762 & 0.3468 & 1.3979 & 0.6238\\
& C6* & 0.2348 & 0.1842 & 2.5169 & 1.1725\\
& C7* & 0.1440 & 0.1223  & 3.6378 & 1.6731\\
\hline
Band 6 & C1 & 1.6215 & 1.3912 & 0.1607 & 0.1071\\
& C2 & 1.2330 & 1.0571 & 0.3137 & 0.1438\\
& C3 & 0.8247& 0.6926 & 0.5002 & 0.2354\\
& C4* & 0.5140 & 0.4651  &0.7835 & 0.3692\\
& C5* & 0.3129 & 0.2873 &1.3979 & 0.6238\\
& C6* & 0.1983 & 0.1537 & 2.5169 & 1.1725\\
& C7* & 0.1186 & 0.1024  &3.6378 & 1.6731\\
\hline
Band 7 & C1 & 1.1427 & 0.9936 & 0.1607 & 0.1071\\
& C2 & 0.8889 & 0.7535 & 0.3137 & 0.1438\\
& C3* & 0.5881 & 0.4968 & 0.5002 & 0.2354\\
& C4* & 0.3598 & 0.3250  &0.7835 & 0.3692\\
& C5* & 0.2179 & 0.2008 &1.3979 & 0.6238\\
& C6* & 0.1371 & 0.1057 &2.5169 & 1.1725\\
& C7* & 0.0823 & 0.0713 & 3.6378 & 1.6731\\
\hline
Band 8$\dagger$ & C1* & 0.9978 & 0.8685 & 0.1607 & 0.1071\\
& C2* & 0.7709 & 0.6477 & 0.3137 & 0.1438\\
& C3* & 0.5030 & 0.4256 & 0.5002 & 0.2354\\
& C4* & 0.3094 & 0.2811  &1.3979 & 0.6238\\
& C5* & 0.1874 & 0.1716 &0.7835 & 0.3692\\
& C6* & 0.1165 & 0.0890 &2.5169 & 1.1725\\
\hline
Band 9$\dagger$ & C1* & 0.6178 & 0.5383 &  0.7835 & 0.3692\\
& C2* & 0.4761 & 0.4002 &  1.3979 & 0.6238\\
& C3* & 0.3091 & 0.2617 &  0.5002 & 0.2354\\
& C4* & 0.1874 & 0.1705  & 0.3137 & 0.1438\\
& C5* & 0.1137 & 0.1034 & 0.1607 & 0.1071\\
\hline
Band 10$\dagger$ & C1* & 0.4779 & 0.4168 &  0.7835 & 0.3692\\
& C2* & 0.3730 & 0.3112 &  1.3979 & 0.6238\\
& C3* & 0.2397 & 0.2035 &  0.5002 & 0.2354\\
& C4* & 0.1463 & 0.1312  & 0.3137 & 0.1438\\
& C5* & 0.0887 & 0.0799 & 0.1607 & 0.1071\\
\hline
\end{tabular}
\end{table*}

\subsection{Millimeter maps at the ALMA receiver bands}
\label{sec:appendix - FOV mm maps}

\begin{figure*}[tbh]
\includegraphics[width=\textwidth]{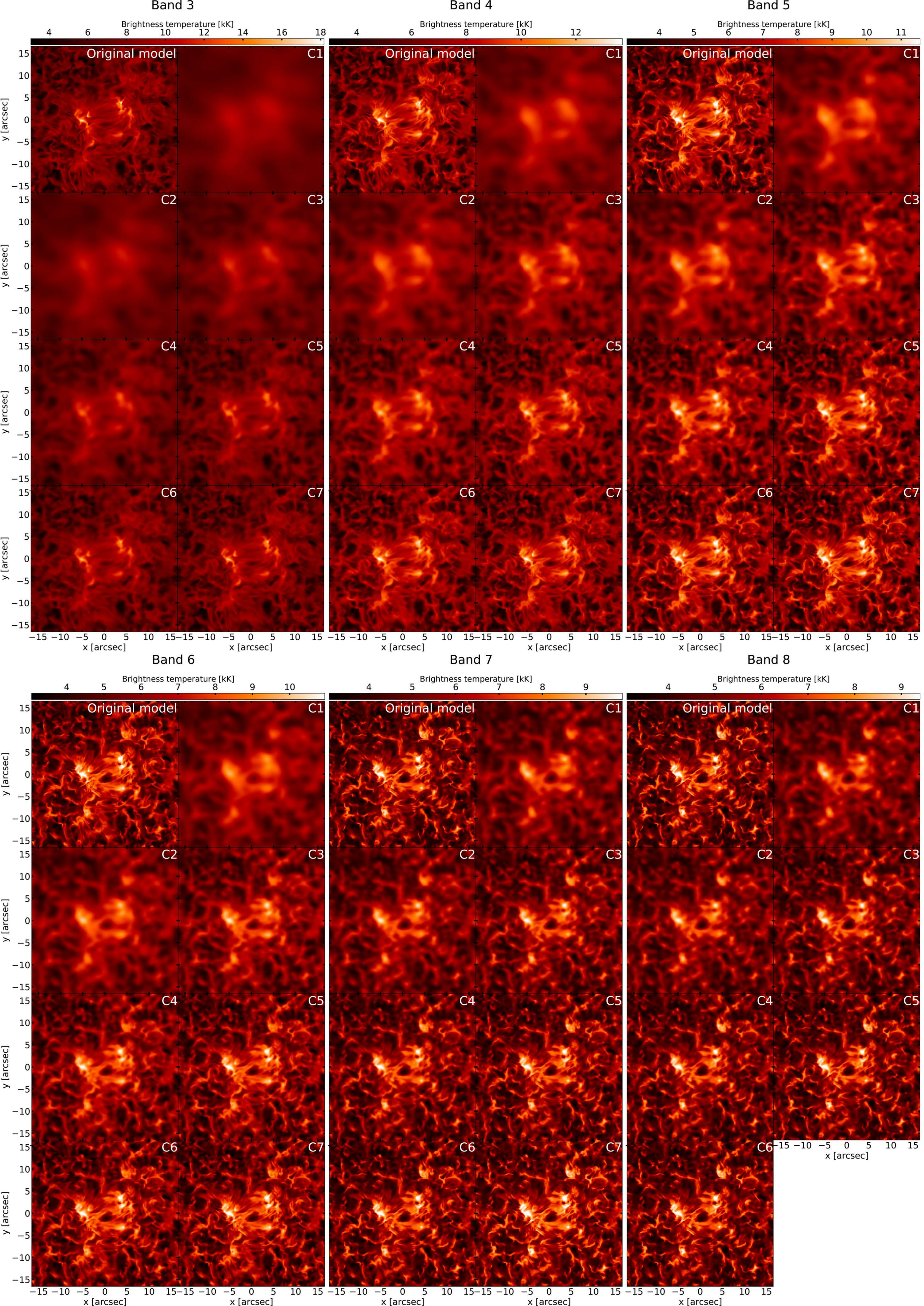}
\caption{FOV of observables at spectral bands $3$--$8$ and at spatial resolutions corresponding to array configurations C1-C7 (up to C6 for band~$8$; \hea{Table~\ref{tab:clean_beam_parameters}}). The color code of each band spans the minimum and $0.9$ times the maximum values of the respective FOV at the original resolution.}
\label{fig:FOV_all_bands3-8}
\end{figure*}

\begin{figure*}[tbh]
\includegraphics[width=\textwidth]{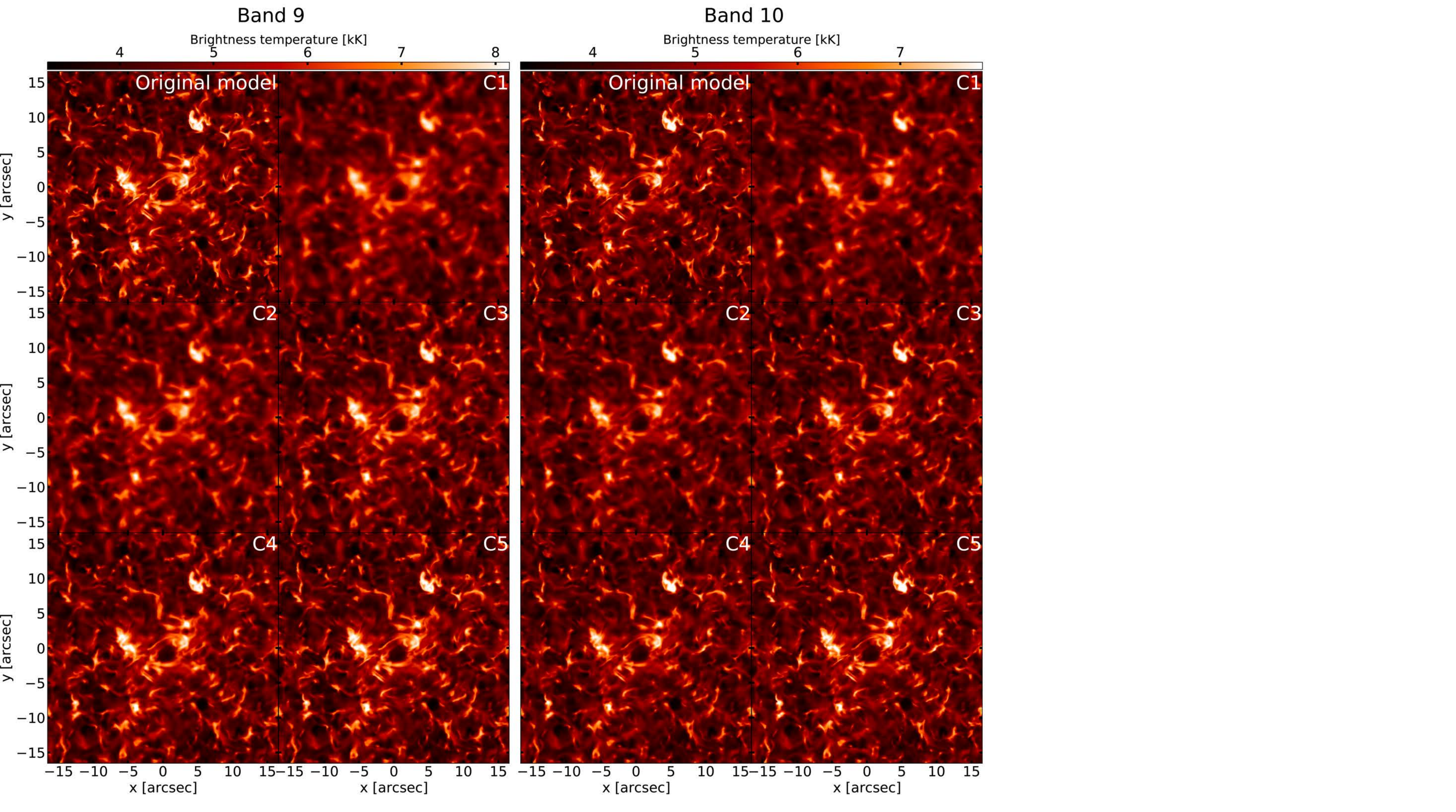}
\caption{FOV of observables at spectral bands $9$--$10$ and at spatial resolutions corresponding to array configurations C1-C5 (\hea{Table~\ref{tab:clean_beam_parameters}}). The color code of each band spans the minimum and $0.9$ times the maximum values of the respective FOV at the original resolution.}
\label{fig:FOV_all_bands9-10}
\end{figure*}

Observables at the wavelengths of each spectral ALMA band are presented at the original resolution of the model and at the degraded resolutions corresponding to the the resolution of the most compact interferometric array configuration C1 to increasingly more extended configurations up to C7 (cf. Table.~\ref{tab:clean_beam_parameters}) for each receiver band in Figs~\ref{fig:FOV_all_bands3-8}~--~\ref{fig:FOV_all_bands9-10}. The mm maps are given for a single snapshot ($t=1650$~s), and the color scales are individually given for each band by the span from the minimum brightness temperature to $0.9$ times the maximum brightness temperature of the FOV at the original resolution. 

\subsection{Degradation of $T_\mathrm{b}$ excess signatures with limited spatial resolution}
\label{sec:appendix_degradation_table}
In this section, we discuss the density distribution plots of the magnitude of Tb excess signatures at original resolution of the model versus degraded resolutions corresponding to the resulting resolutions of ALMA observations at different spectral bands and with different interferometric array configurations (Figs.~\ref{fig:density plots b3 b4}--\ref{fig:density plots b9 b10}). 
\hec{The three targets with different characteristic magnetic field topology (QS, FP and FS) are considered} (Fig.~\ref{fig:FOV}).
Given a spectral band, by direct comparison of the distribution plots at different resolutions in Fig.~\ref{fig:density plots b3 b4}, it becomes clear that a certain spatial resolution is required to resolve brightening events. At low resolution, the distributions are severely squeezed toward the left and show only small magnitudes that are far from the one-to-one ratio. 
These plots naturally contribute with slightly more information than the estimates provided by the linear fits in the main text, and  we gain a more refined view of the ability to resolve the small-scale brightening events and the uncertainties attached to the resulting values with a specific spectral band, target area, and spatial resolution in mind. 
Specifically, valuable information here is the span in parameter space of the remaining fraction of events, outside the $\pm 1 \sigma$ from the median values, and to which degree the signatures at degraded spatial resolution are represented by apparent brightening events. This means that the cases in which the locations of the brightest events in the degraded data do not show bright events at the original resolution (shown in the lower right corner, below the one-to-one ratio in the graphs).

\begin{figure*}[tbh]
\includegraphics[width=\textwidth]{figure_a3.pdf}
\caption{Bands 3 and 4. Density plots of brightening events at the original resolution vs. degraded resolution. Columns show spectral band and target region from left to right: band~3 QS, FS, FP, and band~4 QS, FS and FP. Rows from top to bottom show the resolutions of array configurations C1 -- C7.}
\label{fig:density plots b3 b4}
\end{figure*}

\begin{figure*}[tbh]
\includegraphics[width=\textwidth]{figure_a4.pdf}
\caption{Bands 5 and 6. Density plots of brightening events at the original resolution vs. degraded resolution. Columns show spectral band and target region from left to right: band~5 QS, FS, FP, and band~6 QS, FS and FP. Rows from top to bottom show the resolutions of array configurations C1 -- C7.}
\label{fig:density plots b5 b6}
\end{figure*}

\begin{figure*}[tbh]
\includegraphics[width=\textwidth]{figure_a5.pdf}
\caption{Bands 7 and 8. Density plots of brightening events at the original resolution vs. degraded resolution. Columns show spectral band and target region from left to right: band~7 QS, FS, FP, and band~8 QS, FS and FP. Rows from top to bottom show the resolutions of array configurations C1 -- C7 (limited to C6 for band~8).}
\label{fig:density plots b7 b8}
\end{figure*}

\begin{figure*}[tbh]
\includegraphics[width=\textwidth]{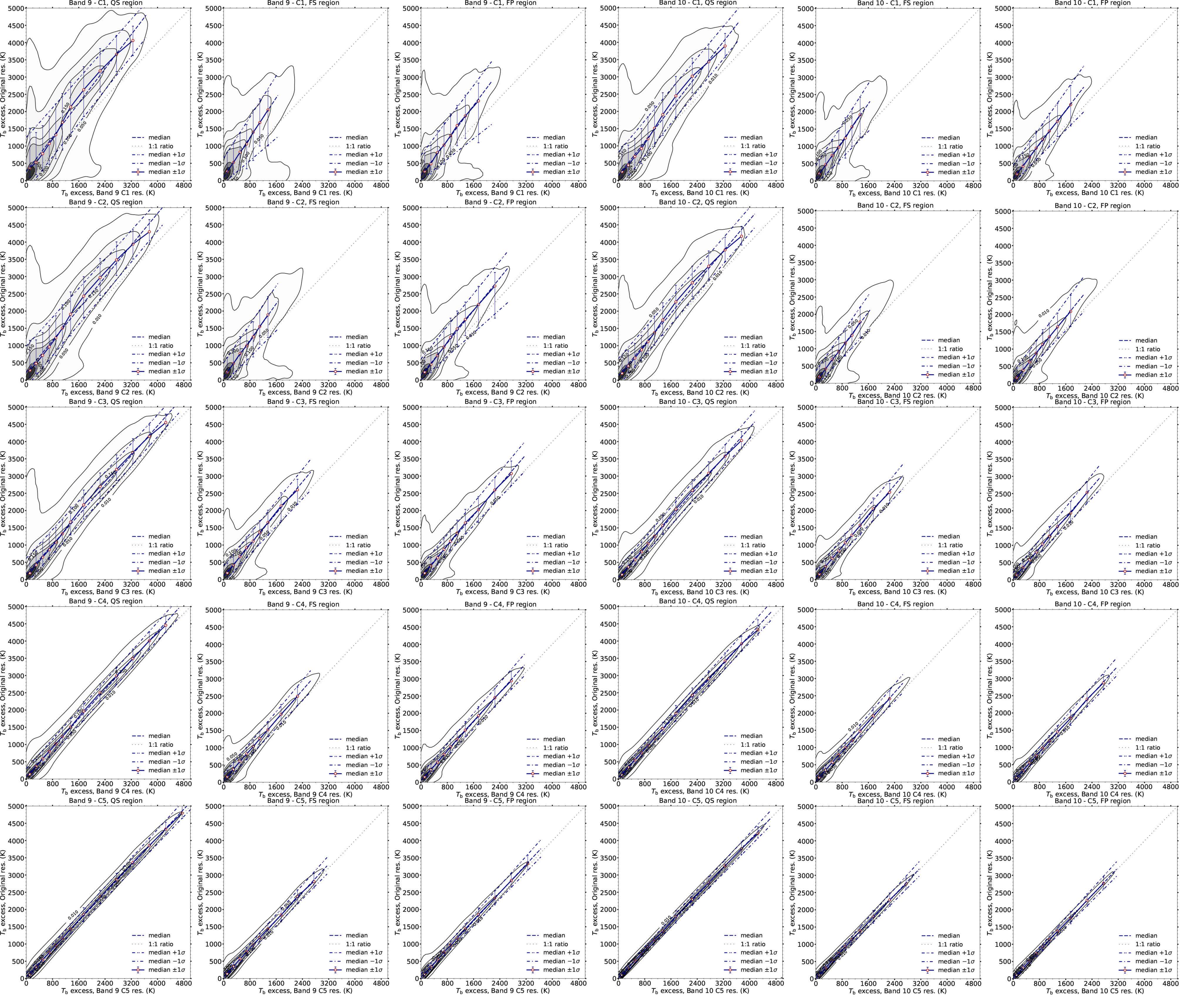}
\caption{Bands 9 and 10. Density plots of brightening events at the original resolution vs. degraded resolution. Columns show spectral band and target region from left to right: band~9 QS, FS, FP, and band~10 QS, FS and FP. Rows from top to bottom show the resolutions of array configurations C1 -- C5.}
\label{fig:density plots b9 b10}
\end{figure*}

\end{document}